\documentclass[12pt]{article}
\usepackage{epsfig}
\usepackage{amssymb,amsmath}
\usepackage{setspace}
\usepackage{subfigure}
\usepackage{amssymb,amsmath}
\usepackage{graphicx}
\usepackage{slashed} 
\usepackage{color}
\usepackage{cancel}
\usepackage[colorlinks=true
,urlcolor=blue
,citecolor=blue
,linkcolor=blue
,pagecolor=blue
,linktocpage=true
,pdfproducer=medialab
]{hyperref}
\def\bi{\begin{itemize}}
\def\ei{\end{itemize}}

\def\to{\rightarrow}

\def\tl{\tilde l}

\def\tst{\tilde t}
\def\ttau{\tilde \tau}

\def\alt{\lesssim}
\def\agt{\gtrsim}

\def\missET{\slashed E_\mathrm{T}} 

\def\mT{m_\mathrm{T}}
\def\mTT{m_\mathrm{T2}}

\def\chia{\tilde{\chi}_1^0}
\def\chib{\tilde{\chi}_2^0}

\def\chiapm{\tilde{\chi}_1^{\pm}}

\def\staua{\tilde{\tau}_1}
\def\GeV{\mathrm{GeV}}

\newcommand\prd[3]{{\it Phys.\ Rev.\ }{\bf D #1} (#2) #3}
\newcommand\prl[3]{{\it Phys.\ Rev.\ Lett.\ }{\bf #1} (#2) #3}

\def\tl{\tilde l}

\def\tst{\tilde t}
\def\ttau{\tilde \tau}

\def\tl{\tilde}

\usepackage{amsmath}

\newcommand{\bea}{\begin{eqnarray}}
\newcommand{\eea}{\end{eqnarray}}

\newcommand{\beq}{\begin{equation}}
\newcommand{\eeq}{\end{equation}}

 \makeatletter
\def\alt{\mathrel{\mathpalette\gl@align<}}
\def\agt{\mathrel{\mathpalette\gl@align>}}
\def\gl@align#1#2{\lower.6ex\vbox{\baselineskip\z@skip\lineskip\z@
\ialign{$\m@th#1\hfil##\hfil$\crcr#2\crcr\sim\crcr}}} \makeatother

\usepackage{longtable}
\graphicspath{{figs/}}

\begin{document}
%
\vspace*{1.0cm}

\begin{center}
\baselineskip 20pt {\Large\bf
Status of Natural Supersymmetry from the GmSUGRA in Light of the current LHC Run-2 and LUX data
}
\vspace{1cm}

\large{
Waqas~Ahmed$^{a,c}$, 
Xiao-Jun Bi$^{b,c}$, 
Tianjun~Li$^{a,c}$, 
Jia Shu Niu$^{a,c}$, 
Shabbar~Raza$^{d}$, 
Qian-Fei Xiang$^{b,c}$,
Peng-Fei Yin$^{b}$
}
\\
\vspace*{0.50cm}
\small{
{\it $^a$ CAS Key Laboratory of Theoretical Physics, Institute of Theoretical Physics, 
Chinese Academy of Sciences, Beijing 100190, China
}
\\
\vspace*{0.2cm}
{\it $^b$ Key Laboratory of Particle Astrophysics, Institute of High Energy Physics, Chinese Academy of Sciences,
Beijing  100049,  China
}\\
\vspace*{0.2cm}
{\it $^c$ School of Physical Sciences, University of Chinese Academy of Sciences,
No.~19A Yuquan Road, Beijing 100049, China}\\
\vspace*{0.2cm}
{\it $^d$
Department of Physics, Federal Urdu University of Arts, Science and Technology, \\
Karachi 75300, Pakistan}\\

}

\vspace{1.25cm} {\bf Abstract}
\end{center}

We study natural supersymmetry in the Generalized Minimal Supergravity (GmSUGRA). For the parameter space 
with low energy electroweak fine-tuning measures less than 50, we are left with only
the $Z$-pole, Higgs-pole and Higgsino LSP scenarios for dark matter (DM). We perform the focused scans 
for such parameter space and find that it satisfies 
various phenomenological constraints and is compatible with the current direct detection bound on neutralino 
DM reported by the LUX experiment. Such parameter space also has solutions with correct DM relic density besides 
the solutions with DM relic density smaller or larger than 5$\sigma$ WMAP9 bounds. We present five benchmark points 
as examples. In these benchmark points, gluino and the first two generations of squarks are heavier than 2 TeV,
stop $\tilde t_{1,2}$ are in the mass range $[1,2]$ TeV, while sleptons are lighter than 1 TeV. Some part of 
the parameter space can explain the muon anomalous magnetic moment within 3$\sigma$ as well. We also perform 
the collider study of such solutions by implementing and comparing with relevant studies done by the ATLAS 
and CMS Collaborations. We find that the points with Higgsino dominant $\chib/\chiapm$ mass up to $300~\GeV$ 
are excluded in $Z$-pole scenario while for Higgs-pole scenario, the points with $\chib$ mass up to $460~\GeV$ are excluded. We also notice that the Higgsino LSP points in our present scans are beyond the reach of present LHC searches. 
Next, we show that for both the $Z$-pole and Higgs-pole scenarios, the points with electroweak fine-tuning measure 
around 20 do still survive.

\thispagestyle{empty}

\newpage

\addtocounter{page}{-1}

\baselineskip 18pt

\section{Introduction}

Undoubtedly, the gauge coupling unification of the strong, weak and electromagnetic interactions of 
the fundamental particles is a great triumph of the supersymmetric (SUSY) version of the Standard Model (SM) 
of particle physics~\cite{gaugeunification},  which henceforth will be called as Supersymetric SM (SSM). 
The SSM predicts the existence of SUSY partners of all the known SM particles. Interestingly, the existance 
of these particles can help us to understand the stabilization of the electroweak (EW) scale and thus 
solves yet another daunting problem of particle physics named as the gauge hierarchy problem~\cite{ghp}. 
In addition, the Minimal SSM (MSSM) also predicts the Higgs boson mass ($m_{h}$) should be smaller 
than 135 GeV~\cite{mhiggs}. Indeed, the ATLAS and CMS Collaborations of the Large Hadron Collider (LHC) 
have discovered a SM-like Higgs boson $h$ with mass $m_{h}$= 125 GeV~\cite{Aad:2012tfa,CMS}. This adds 
yet another feather in the hat of the SSM. The SSM also predicts that with $R$-parity conservation, 
the Lightest Supersymmetric Particle (LSP) such as neutralino is an excellent dark matter 
candidate \cite{neutralinodarkmatter,darkmatterreviews}.  And the electroweak symmetry can be broken 
radiatively due to large top quark Yukawa coupling, etc. All these observations give us some hints 
that we are on the right track. 

The existence of the SM-like Higgs boson with mass $m_{h}\sim$ 125 GeV requires the multi-TeV top squarks with
small mixing or TeV-scale top squarks with large mixing. This raises a question on the naturalness of the MSSM 
and generates the fine-tuning problem. However, the null results of the LHC-Run2 and the ongoing LHC SUSY-searches 
have not found any SUSY evidences yet. In recent studies, the bounds on squark masses $m_{\tilde q} \gtrsim$ 1600 GeV~\cite{lhc-strong} 
and gluino mass $m_{\tilde g}\gtrsim$ 2000 GeV \cite{lhc-strong} have been reported by the ATLAS and CMS Collaborations 
at the 13 TeV LHC with $36~\mathrm{fb}^{-1}$ of data. This situation has put the promises of the MSSM under pressure. 
It is interesting to note that despite the SM-like Higgs mass being relatively heavy,  
there are some studies~\cite{Drees:2015aeo,Ding:2015epa,Baer:2015rja,Batell:2015fma,AbdusSalam:2015uba,Barducci:2015ffa,Cohen:2015ala,Fan:2014axa,Leggett:2014hha,Dimopoulos:2014aua,Gogoladze:2013wva,Kribs:2013lua,Gogoladze:2012yf}
which suggest that the naturalness problem in the MSSM can be solved successfully. In particular, 
in an interesting scenario, which is called as Super-Natural SUSY~\cite{Leggett:2014hha, Du:2015una}, it can be shown 
that no residual electroweak fine-tuning  (EWFT) left in the MSSM if we employ the No-Scale supergravity boundary 
conditions~\cite{Cremmer:1983bf} and Giudice-Masiero (GM) mechanism~\cite{Giudice:1988yz} despite having   
relatively heavy spectra. Some people might think that the Super-Natural SUSY might have a problem related to 
the higgsino mass parameter $\mu$, which is generated by the GM mechanism and is proportional to the universal gaugino mass $M_{1/2}$,
since the ratio $M_{1/2}/\mu$ is of order one but cannot be determined as an exact number.
 This problem, if it is,  can be addressed in the M-theory inspired the Next to 
MSSM (NMSSM) \cite{Li:2015dil}. Also, see~\cite{Baer:2017pba}, for more recent works related to naturalness within and beyond the MSSM. 

In order to quantify the amount of fine-tuning (FT), we need to define the fine-tuning measures.
In literatures, we can find the high energy fine-tuning measure $\Delta_{EENZ-BG}$ defined by 
Ellis, Enqvist, Nanopoulos and Zwirner~\cite{EENZ}, as well as Barbieri and Giudice~\cite{Barbieri:1987fn}, 
and the high energy and electroweak fine-tuning measures 
$\Delta_{HS}$ and $\Delta_{EW}$ defined by Baer, Barger, Huang, Michelson, Mustafayev and Tata~\cite{Baer:2012up,Baer:2012mv}.  
Usually, we have $\Delta_{EW} \lesssim \Delta_{BG} \lesssim \Delta_{HS}$. One can show that 
$\Delta_{EW}\sim \Delta_{BG}$ for some scenarios~\cite{Ahmed:2016lkh}.

This work is a continuation of our phenomenological studies of Generalized Minimal Supergravity Model (GmSUGRA)~\cite{Li:2010xr}. 
In Refs.~\cite{Li:2014dna,Li:2016ucz}, we showed that in GmSUGRA, we have varieties of dark matter scenarios such as 
$A$-resonance, Higgs-resonance, $Z$-resonance, stau-neutralino coannihilation, tau sneutrino-neutralino coannihilation 
compatible with various phenomenological constraints. In addition, we showed that the Higgs coupling and muon anomalous magnetic moment measurements can constrain the parameter space effectively. In this work, 
we concentrate on the dark matter solutions which not only have 
low EWFT (that is $\Delta_{EW} \lesssim 50$), but also are consistent with current direct detection bounds 
reported by the LUX Collaboration ~\cite{Akerib:2016vxi}. In our scans, we find that the light  
stau-neutralino coannihilation 
points do not satisfy $\Delta_{EW}\lesssim$ 50. Also, the Higgsino LSP points are still natural and viable, 
 but they cannot be probed at the current LHC searches. 
We find that only Higgs-pole and Z-pole solutions fulfil the above mentioned criteria. Therefore, 
we will only consider these two type of resonance points in more details. In these two scenarios, 
a subset of solutions satisfy the 5$\sigma$ dark matter relic density WMAP9 bounds while the other solutions have relic density beyond the 5$\sigma$ bounds. We present five benchmark points as examples of the parameter space under consideration, where one of them has the Higgsino LSP. In these benchmark points, gluino and 
the first two generations of squarks are heavier than 2 TeV,
top squars $\tilde t_{1,2}$ are in the mass range $[1,2]$ GeV, while sleptons are less than 1 TeV. 
Some part of the parameter space can also explain the muon anomalous magnetic moment 
within 3$\sigma$~\cite{Bennett:2006fi}. Furthermore, we consider the constraints on such solutions from the direct searches 
for the SUSY particles at the LHC. In order to realize small fine-tuning and satisfy experimental constraints simultaneously, 
only electroweakinos (neutralinos and charginos) and stau are light and could be explored at the
current LHC searches. We study various electroweak Drell-Yan production processes where one 
could  produce neutralinos which could decay through on-shell or off-shell  $Z^{(\ast)}$ ($\tilde{\chi}_i^0 \to	Z^{(\ast)} \chia$) or $h^{(\ast)}$ ($\tilde{\chi}_i^0 \to h^{(\ast)} \chia$). We will give more details about the our analyses later in this paper.
We display various plots showing that the relevance of different decay modes depends on mass spectra and will significantly influence collider searches for these particles. The dominant decay channel of $\chib$ for samples of $Z$-pole is $\chib \to \chia Z^{(\ast)}$ when the mass difference $m_{\chib}-m_{\chia}$ is small.
Once the decay into Higgs boson is kinematically possible, branching ratio to $\chia h$ increase with increasing of $m_{\chib}-m_{\chia}$ and become the dominant channel when $m_{\chib}-m_{\chia} \gtrsim 140~\GeV$. The decay channels of $\chiapm$ is always $\chiapm \to \chia W^{(\ast)}$. We also find that for our present work, $3l+\missET$ and $2l+\missET$ give the best sensitivity at the LHC searches where electroweakinos decays to multi-leptons. We use suitable kinematic variables to discriminate signals from backgrounds. We show the $95\%$ C.L. exclusion results of the LHC electroweakinos searches in the $m_{\chia}$-$m_{\chib}$ plane and $m_{\chia}$-$\Delta_{\mathrm{EW}}$ plane. It can be seen from these plots that higgsino dominant $\chib/\chiapm$ with mass up to $300~\GeV$ are excluded in case of $Z$-pole while for Higgs-pole scenario, points with $\chib$ mass up to $460~\GeV$ are excluded. Moreover, it can also be noticed that $Z$-pole solutions with small $\Delta_{\mathrm{EW}}$ are easy to be explored, whereas solutions with large $\Delta_{\mathrm{EW}}$ are hard to exclude but for the Higgs-pole, many points with $\Delta_{\mathrm{EW}}$ up to 50 could by excluded by electroweakino searches with tau final states. Finally, we notice that for both the $Z$-pole and the Higgs-pole, samples with $\Delta_{\mathrm{EW}} \sim$ 20  could still survive, indicating naturalness of this SUSY framework.  

The remainder of this paper is organized as follows. We present our model in Section~\ref{model}. We discuss EWFT measure in Section~\ref{ewft}. Section~\ref{sppc} is devoted for scanning procedure and phenomenological constraints.
Our results for focused scans are shown in Section~\ref{fs} while results for the LHC searches are presented in Section~\ref{lhc_searches}. A summary and conclusion are given in Section~\ref{dc}.

\section{The Electroweak SUSY from the GmSUGRA in the MSSM}
\label{model}
In GmSUGRA, at the GUT-scale, we can write the generalized gauge coupling relation and 
the generalized gaugino mass relation as follows
\begin{eqnarray}
{{1}\over {\alpha_2}} - {{1}\over {\alpha_3}} 
~=~k \left( {{1}\over {\alpha_1}} 
- {{1}\over {\alpha_3}} \right) ~,~\,
\label{GCRelation}
\end{eqnarray}
\begin{eqnarray}
{{M_2}\over {\alpha_2}} - {{M_3}\over {\alpha_3}} 
~=~k \left( {{M_1}\over {\alpha_1}} 
- {{M_3}\over {\alpha_3}} \right) ~,~\,
\label{GMRelation}
\end{eqnarray}
where $k$ is the index of these relations since it is invariant 
under one-loop Renormalization Group Equation (RGE) running. 
For more details about the model, please see~\cite{Li:2010xr}.

Another important feature of GmSUGRA is that we can realize Electroweak SUSY (EWSUSY). 
In this scenario,
we can have the sleptons and electroweakinos within one TeV while squarks and/or gluinos 
can be in several TeV mass ranges~\cite{Cheng:2012np}. Assuming gauge coupling unification at the GUT scale ($\alpha_1=\alpha_2=\alpha_3$) and using $k=5/3$, we obtain a simple gaugino mass relation from Eq.~(\ref{GMRelation})
\begin{equation}
 M_2-M_3 = \frac{5}{3}~(M_1-M_3)~.
\label{M3a}
\end{equation}
It is straightforward to notice that
the universal gaugino mass relation $M_1 = M_2 = M_3$ in the mSUGRA, is just a special case of 
this general one. This is why we call it Generalized mSUGRA. We will choose $M_1$ and $M_2$ to be free input parameters, 
which vary around several hundred GeV for the EWSUSY. We can now write Eq.~(\ref{M3a}) for $M_3$ as:
\begin{eqnarray}
M_3=\frac{5}{2}~M_1-\frac{3}{2}~M_2~,
\label{M3}
\end{eqnarray}
which could be as large as several TeV or as small as several hundred GeV, depending
 on specific values of $M_1$ and $M_2$.

The general SUSY breaking (SSB) soft scalar masses at the GUT scale are given 
in Ref.~\cite{Balazs:2010ha}. 
Taking the slepton masses as free parameters, we obtain the following squark masses 
in the $SU(5)$ model with an adjoint Higgs field
\begin{eqnarray}
m_{\tl{Q}_i}^2 &=& \frac{5}{6} (m_0^{U})^2 +  \frac{1}{6} m_{\tl{E}_i^c}^2~,\\
m_{\tl{U}_i^c}^2 &=& \frac{5}{3}(m_0^{U})^2 -\frac{2}{3} m_{\tl{E}_i^c}^2~,\\
m_{\tl{D}_i^c}^2 &=& \frac{5}{3}(m_0^{U})^2 -\frac{2}{3} m_{\tl{L}_i}^2~,
\label{squarks_masses}
\end{eqnarray}
where $m_{\tl Q}$, $m_{\tl U^c}$, $m_{\tl D^c}$, $m_{\tl L}$, and  $m_{\tl E^c}$ represent the scalar masses of
the left-handed squark doublets, right-handed up-type squarks, right-handed down-type squarks,
left-handed sleptons, and right-handed sleptons, respectively, while $m_0^U$ is the universal  
scalar mass, as in the mSUGRA. In the EWSUSY, $m_{\tl L}$ and $m_{\tl E^c}$ are both within 1 TeV, resulting in 
light sleptons. Especially, in the limit $m_0^U \gg m_{\tl L/\tl E^c}$, we have the approximated 
relations for squark masses: $2 m_{\tl Q}^2 \sim m_{\tl U^c}^2 \sim m_{\tl D^c}^2$. In addition, 
the Higgs soft masses $m_{\tl H_u}$ and $m_{\tl H_d}$, and the  trilinear soft terms
 $A_U$, $A_D$ and $A_E$ can all be free parameters from the GmSUGRA~\cite{Cheng:2012np, Balazs:2010ha}.

\section{The Electroweak Fine Tuning}\label{ewft}
As we mentioned earlier that in this work we are interested in solutions with low EWFT. We use the (7.85) version of  ISAJET \cite{ISAJET} to calculate the  FT conditions 
at the EW scale $M_{EW}$. After including the one-loop effective potential contributions to the tree-level MSSM Higgs potential, the $Z$-boson mass $M_Z$ is given by
\begin{equation}
\frac{M_Z^2}{2} =
\frac{(m_{H_d}^2+\Sigma_d^d)-(m_{H_u}^2+\Sigma_u^u)\tan^2\beta}{\tan^2\beta
-1} -\mu^2 \; ,
\label{eq:mssmmu}
\end{equation}
where $\Sigma_u^u$ and  $\Sigma_d^d$ are the contributions coming from the one-loop effective potential 
defined in Ref.~\cite{Baer:2012mv} and $\tan\beta \equiv \frac{v_u}{v_d}$. All parameters  
in Eq. (\ref{eq:mssmmu}) are defined at the $M_{EW}$.
In order to measure the EWFT condition we follow \cite{Baer:2012mv} and use the following definitions
\begin{equation}
 C_{H_d}\equiv |m_{H_d}^2/(\tan^2\beta -1)|,\,\, C_{H_u}\equiv
|-m_{H_u}^2\tan^2\beta /(\tan^2\beta -1)|, \, \, C_\mu\equiv |-\mu^2 |,
\label{cc1}
\end{equation}
 with each $C_{\Sigma_{u,d}^{u,d} (r)}$  less than some characteristic value of order $M_Z^2$.
Here, $r$ labels the SM and SUSY particles that contribute to the one-loop Higgs potential.
For the fine-tuning measure we define
\begin{equation}
 \Delta_{\rm EW}\equiv {\rm max}(C_r )/(M_Z^2/2)~.
\label{eq:ewft}
\end{equation}
Note that $\Delta_{EW}$ only depends on the weak-scale parameters of the SSMs, and then is fixed
by the particle spectra. Hence, it is independent of how the SUSY particle masses arise. 
Lower values of $\Delta_{EW}$
corresponds to less fine tuning, for example, $\Delta_{EW}=50$ implies $\Delta_{EW}^{-1}=2\%$ fine tuning.
In addition to $\Delta_{EW}$, ISAJET also calculates $\Delta_{HS}$ which is a measure of fine-tuning at 
the High Scale (HS) like the GUT scale
in our model~\cite{Baer:2012mv}. The HS fine-tuning measure $\Delta_{HS}$ is given as follows
\begin{equation}
 \Delta_{\rm HS}\equiv {\rm max}(B_i )/(M_Z^2/2)~.
\label{eq:hsft}
\end{equation}
For definition of $B_i$ and more details, please see Ref.~\cite{Baer:2012mv}. 

\section{Scanning Procedure and Phenomenological Constraints }
\label{sppc}
We  employ the ISAJET~7.85 package~\cite{ISAJET}
 to perform the 
focused scans using parameters given in Section~\ref{model} to explore the parameter space having 
$Z$-resonance and Higgs-resonance solutions. In this work, we will focus on the solutions 
with relatively small EWFT $\Delta_{EW} \lesssim$ 50. For full ranges of the parameter see \cite{Li:2014dna}.  

In ISAJET, the weak scale values of the gauge and third
 generation Yukawa couplings are evolved to
 $M_{\rm GUT}$ via the MSSM renormalization group equations (RGEs)
 in the $\overline{DR}$ regularization scheme.
We do not strictly enforce the unification condition
 $g_3=g_1=g_2$ at $M_{\rm GUT}$, since a few percent deviation
 from unification can be assigned to the unknown GUT-scale threshold
 corrections~\cite{Hisano:1992jj}.
With the boundary conditions given at $M_{\rm GUT}$,
 all the SSB parameters, along with the gauge and Yukawa couplings,
 are evolved back to the weak scale $M_{\rm Z}$.

In evaluating Yukawa couplings, the SUSY threshold
 corrections~\cite{Pierce:1996zz} are taken into account
 at the common scale $M_{\rm SUSY}= \sqrt{m_{\tst_L}m_{\tst_R}}$.
The entire parameter set is iteratively run between
 $M_{\rm Z}$ and $M_{\rm GUT}$ using the full two-loop RGEs
 until a stable solution is obtained.
To better account for the leading-log corrections, one-loop step-beta
 functions are adopted for gauge and Yukawa couplings, and
 the SSB parameters $m_i$ are extracted from RGEs at appropriate scales
 $m_i=m_i(m_i)$.
The RGE-improved one-loop effective potential is minimized
 at an optimized scale $M_{\rm SUSY}$, which effectively
 accounts for the leading two-loop corrections.
The full one-loop radiative corrections are incorporated
 for all sparticles.

It should be noted that the requirement of radiative electroweak symmetry breaking (REWSB)~\cite{Ibanez:1982fr} puts an important theoretical
 constraint on parameter space.
Another important constraint comes from limits on the cosmological
 abundance of stable charged particle~\cite{Beringer:1900zz}.
This excludes regions in the parameter space where charged
 SUSY particles, such as $\ttau_1$ or $\tst_1$,
 become the LSP.
We accept only those solutions for which one of the neutralinos
 is the LSP.

Also, we consider  $\mu > 0$ and  use $m_t = 173.3\, {\rm GeV}$  \cite{:2009ec}.
Note that our results are not too sensitive to one
 or two sigma variations in the value of $m_t$  \cite{bartol2}.
We use $m_b^{\overline{DR}}(M_{\rm Z})=2.83$ GeV as well
 which is hard-coded into ISAJET.  Also, we will use the notations $A_t,~A_b,~A_{\tau}$ 
for $A_{U},~A_D$ and $A_E$, receptively.


In scanning the parameter space, we employ the Metropolis-Hastings
 algorithm as described in \cite{Belanger:2009ti}.
The data points collected all satisfy the requirement of REWSB,
 with the neutralino being the LSP.
After collecting the data, we require the following bounds (inspired by the LEP2 experiment) 
on sparticle masses.\\
\noindent
\textbf{({1}) LEP2 constraints}

We employ the LEP2 bounds on sparticle masses 
\begin{eqnarray}\label{eqn:spMassLEP2}
m_{\tilde{t}_1}, m_{\tilde{b}_1} & \geq & 100\, {\rm GeV}, \nonumber \\
m_{\tilde{\tau}_1} & \geq & 105\, {\rm GeV}, \nonumber \\
m_{\tilde{\chi}_1^{\pm}} & \geq & 103\, {\rm GeV}.
\end{eqnarray}

\noindent
\textbf{({2}) Higgs mass constraints}

The combined value of Higgs mass reported by the ATLAS and CMS Collaborations is \cite{Khachatryan:2016vau}
\begin{eqnarray}
m_{h} = 125.09 \pm 0.21(\rm stat.) \pm 0.11(\rm syst.)~ GeV.
\end{eqnarray}  
Due to the theoretical uncertainty in the Higgs mass calculations in the MSSM \cite{Allanach:2004rh}, 
we use the following Higgs mass bound  
\begin{eqnarray}\label{eqn:higgsMassLHC}
122\, {\rm GeV} \leq m_h \leq 128\, {\rm GeV}\, . 
\end{eqnarray}

\noindent
\textbf{({3}) LHC constraints}

We demand~\cite{lhc-strong}
\begin{eqnarray}
m_{\tilde q} \geq 2000\, {\rm GeV}\,~,~ \\ \nonumber 
m_{\tilde g} \geq 2000\, {\rm GeV}~.~
\end{eqnarray}

\noindent
\textbf{({4}) B-physics constraints}

We use the IsaTools package \cite{bsg, bmm} and implement the following B-physics constraints
\begin{eqnarray}\label{eqn:Bphysics}
1.6\times 10^{-9} \leq{\rm BR}(B_s \rightarrow \mu^+ \mu^-) 
  \leq 4.2 \times10^{-9} \;(2\sigma)~~&\cite{CMS:2014xfa} ~,~& 
\\ 
2.99 \times 10^{-4} \leq 
  {\rm BR}(b \rightarrow s \gamma) 
  \leq 3.87 \times 10^{-4} \; (2\sigma)~~&\cite{Amhis:2014hma}~,~ &  
\\
0.70\times 10^{-4} \leq {\rm BR}(B_u\rightarrow\tau \nu_{\tau})
        \leq 1.5 \times 10^{-4} \; (2\sigma)~~&\cite{Amhis:2014hma}~.~&
\end{eqnarray}

\noindent
\textbf{({5}) Electroweak Fine-Tuning constraint}

Because we  consider the natural SUSY, 
the following constraint on fine-tuning measure $\Delta_{\rm EW}$ is applied
\begin{eqnarray}\label{eqn:deltaEW}
\Delta_{\rm EW} \leq 50.
\end{eqnarray}
\noindent

\textbf{({6}) WMAP constraint}

We apply the WMAP9 bounds with 5$\sigma$ variation on DM density~\cite{Bennett:2012zja}
\begin{eqnarray}\label{eqn:wmap_cons}
0.0913 \, \le \Omega h^2 \le 0.1363~.
\end{eqnarray}


\section{Results of focused scans}\label{fs}

\begin{figure}[htp]
\centering
\subfiguretopcaptrue

\subfigure{
\includegraphics[totalheight=5.5cm,width=7.cm]{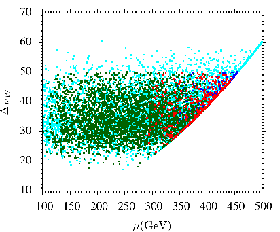}
}
\subfigure{
\includegraphics[totalheight=5.5cm,width=7.cm]{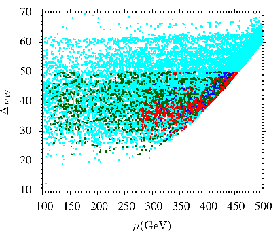}
}
\subfigure{
\includegraphics[totalheight=5.5cm,width=7.cm]{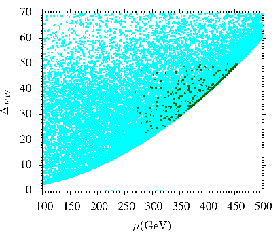}
}
\subfigure{
\includegraphics[totalheight=5.5cm,width=7.cm]{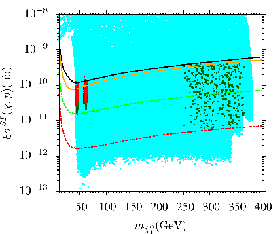}
}
\caption{$\Delta_{EW}$ vs. $\mu$ for the $Z$-pole (left), for Higgs-pole (right) and Higgsino LSP (bottom left) scenario.
Rescaled spin-independent ($\xi \sigma^{SI}(\chi,p)$) rate vs. LSP neutralino mass $m_{\tilde \chi_{1}^{0}}$ (bottom right). 
Aqua points satisfy the REWSB and LSP neutralino  conditions. Red, blue and green solutions  represent 
the sets of points with relic density consistent with, greater than and smaller than 5$\sigma$ WMAP9 bounds, respectively. 
These points also satisfy the bounds indicated in Section~\ref{sppc}.
}
\label{fig1}
\end{figure}

We present results of focused scans in Fig.~\ref{fig1}.
In the top right and left panels we display plots in $\Delta_{EW}$ vs. $\mu$ respectively for $Z$-pole 
and for Higgs-pole scenarios while 
rescaled spin-independent ($\xi \sigma^{SI}(\chi,p)$) rate vs. LSP neutralino mass $m_{\tilde \chi_{1}^{0}}$ is shown in bottom panel. Aqua points satisfy the REWSB and LSP neutralino  conditions. Red, blue and green points represent 
the sets of points respectively with DM relic density consistent with, greater than,
 and smaller than 5$\sigma$ WMAP9 bounds, as well as consistent with upper bounds reported by the LUX experiment. 
These points all also satisfy the bounds given in Section~\ref{sppc}. We see that green points both for $Z$-pole 
and Higgs-pole scenarios have
$\Delta_{EW}$ in the range 20 to 50, while $\mu$ is in the range of $[100,450]$ GeV. Blue and red points have $\Delta_{EW}$ as small as 24 and $\mu$ is confined between $[300,450]$ GeV. In the bottom left panel, we show a plot in the same plane for Higgsino LSP solutions. Because such solutions have small relic densities, all the  points are green. One can see that such solutions have $\Delta_{EW}$ values from 20 to 50 with corresponding $\mu$ values varying roughly from 250 GeV to 450 GeV. We will talk about Higgsino LSP solutions more with reference to the plots in $\xi \sigma^{SI}(\chi,p)$-$m_{\tilde \chi_{1}^{0}}$ plane. 
\begin{table}[htp!]
\centering
\begin{tabular}{|c|ccccc|}
\hline
\hline

                 & BMP 1 & MMP 2 & BMP 3 & BMP 4 & BMP 5\\

\hline
$m_{0}$             & 1449     & 1424     & 1537 & 1367 & 1053\\
$m_{\tilde Q}$      & 1358.3   & 1323.3   &  1404.1   & 1249.4 & 1021.8\\
$m_{\tilde U^c}$    & 1765.8   & 1770.3   &  1981.3  & 1760.5 & 1169.3\\
$m_{\tilde D^c}$    & 1715.7   & 1701.7   &  1945.9   & 1717.4 & 1189.1\\
$m_{\tilde L}$      & 912.9    & 851.8     &  475.7     & 497.6 & 806.8\\
$m_{\tilde E^c}$    & 756.2    & 607       &  132.2     & 151.3 & 849.1\\
$M_{1} $            & 96.81    & 98.19       &  132.6    & 131.1 & 857.7\\
$M_{2}$             & 812.9    & 751.8      &  1023  &  1105 & 706.8\\
$M_{3}$             & -977.33  & -882.22   & -1203  &-1329.8 & 1084.1\\
$A_t=A_b$           & 3632    & 3689       &  4981   & 5076 & 857.7\\
$A_{\tilde \tau}$   & -403.1   & -413.5   &  -238.2  & -186.9 & -2915\\
$\tan\beta$         & 17.6    & 18.9      & 21.3     & 19.8 & 14.5\\
$m_{H_u}$           & 2631   & 2562       & 3231   & 3306 & 2558\\
$m_{H_d}$           & 2618    & 2697      & 3284   & 3203 & 487.6\\
\hline
$\mu$               & 326 & 254  &   351  & 276 & 262\\
$\Delta_{EW}$       & 29  & 24   &   35  & 31 & 29\\
$\Delta_{HS}$       & 1691& 1597 &    2552  & 2660 & 1597\\
$\Delta a_{\mu}$    &$4.17 \times 10^{-10}$ &$5.59 \times 10^{-10}$  &$6.34\times 10^{-10}$  & $ 4.93 \times 10^{-10}$ & $3.61\times 10^{-10}$\\
\hline

\hline
$m_h$            & 123  & 123     & 125  & 125 & 123\\
$m_H$             &2515   & 2553    & 3060  & 3026 & 529\\
$m_A$            & 2499   & 2536     & 3040  & 3006 & 525\\
$m_{H^{\pm}}$    & 2516   & 2554     & 3061  & 3027 & 534\\
\hline
$m_{\tilde{\chi}^0_{1,2}}$
                 & 45.9, 326 & 45, 255     & 62, 355  & 62, 283 & 248, 271\\

$m_{\tilde{\chi}^0_{3,4}}$
                 & 337,712  & 266, 658    & 363, 882   & 287, 953 & 373, 587\\

$m_{\tilde{\chi}^{\pm}_{1,2}}$
                 & 333, 704  & 260, 651    & 362, 876  & 286, 946 & 265, 579\\
\hline
$m_{\tilde{g}}$  & 2220 & 2025  & 2676  & 2918 & 2397\\
\hline
$m_{ \tilde{u}_{L,R}}$
                 & 2374, 2542  & 2216, 2411   & 2752, 2975   & 2873, 3026 &2322, 2421\\
$m_{\tilde{t}_{1,2}}$
                 & 1173, 1731 & 1000, 1542   & 1069, 1811   & 1046, 1960 & 1062, 1760\\
\hline
$m_{ \tilde{d}_{L,R}}$
                 & 2375, 2561 & 2218, 2434    & 2753, 3016  & 2875, 3047 & 2323, 2366\\
$m_{\tilde{b}_{1,2}}$
                 & 1717, 2433   & 1525, 2287   &  1812, 2777   & 1969, 2831 & 1734, 2285\\
\hline
$m_{\tilde{\nu}_{1,2}}$
                 &978 & 878   & 670  & 774 & 1002\\
$m_{\tilde{\nu}_{3}}$
                 &935 & 821    & 532   & 679 & 996\\
\hline
$m_{ \tilde{e}_{L,R}}$
                & 984, 909  & 883, 839    & 683   & 786, 522 & 1008, 729\\
$m_{\tilde{\tau}_{1,2}}$
                & 816, 941   & 719, 828   & 264, 549   & 162, 693 & 716, 1001\\
\hline

$\sigma_{SI}({\rm pb})$
                & $8.05\times 10^{-11}$ & $ 1.6 4\times 10^{-10} $  & $7.33\times 10^{-11}$  & $1.64\times 10^{-10}$ & $  1.71 \times 10^{-8} $\\

$\sigma_{SD}({\rm pb})$
                & $ 1.19\times 10^{-5}$ &$ 3.38 \times 10^{-5} $  & $9.07\times 10^{-6}$ & $2.40\times 10^{-5}$ & $ 1.39 \times 10^{-4} $\\

$\Omega_{CDM}h^{2}$ &0.106  & 0.017  & 0.103  & 0.022 & 0.002 \\
\hline
\hline
\end{tabular}
\caption{All the masses in this table are in units of GeV.
All the points satisfy the constraints  described in Section~\ref{sppc}. BMP1 and BMP2 are examples of 
$Z$-pole solutions, BMP3 and BMP4 are representatives of Higgs-pole solutions,
and BMP 5 is an example for Higgsino LSP solutions.
}
\label{table1}
\end{table}

In $\xi \sigma^{SI}(\chi,p)$-$m_{\tilde \chi_{1}^{0}}$ plot, solid black and red lines respectively 
represent the current LUX \cite{Akerib:2016vxi} and XENON1T \cite{Aprile:2017iyp} bounds. 
The dashed green and red lines display projection of XENON1T for next two years and XENONnT 
(total exposure of 20 t.y) \cite{Aprile:2015uzo}, respectively. The factor $\xi \equiv \Omega h^2 /0.12$
for green points which accounts for a possible depleted local abundance of neutralino DM, 
while $\xi =$ 1 for red and blue points. In this plot, the two dips around 45 GeV and
62 GeV indicate the $Z$-pole and Higgs-pole solutions. Here, we want to make a comment that in
focused scans, we also got points beyond the current LUX bounds but we have chopped them out 
and have displayed throughout this work only those solutions which are consistent with these bounds.
By the way, if we introduce an axino as the LSP, {\it i.e.}, the lightest neutralino is
not the LSP, these chopped points are still natural
and consistent with all the current experimental constraints.
Moreover, we can  see that in the near future the XENON1T experiment will completely probe solutions of our present scans. One can also notice that there is a wide gap between the Higgs-pole solutions and Higgsino LSP solutions (green points with $m_{\tilde \chi_{1}^{0}}$ between 250 GeV 350 GeV). We notice that the $\sigma^{SI}(\chi,p)$ is too high for points with neutralino mass between 65 GeV to 250 GeV. Even if we rescale the $\sigma^{SI}(\chi,p)$, points still rule out by the current LUX bounds. In addition to it, 
 we also notice that for the Higgsino LSP scenario, $\chi_1^0$, $\chi_2^0$, and 
$\chi_1^\pm$ are Higgsino dominated, $\chi_3^0$ is Bino dominated,  $\chi_4^0$ and $\chi_2^\pm$ are Wino dominated.   
Since $m_{\chi_1^0} \sim m_{\chi_2^0} \sim m_{\chi_1^\pm}$, leptons from $\chi_2^0$ and $\chi_1^\pm$ are hard to reconstruct.  
The most effective channels that could contribute to 3 leptons is $p p \rightarrow \chi_4^0 \chi_2^\pm$. However, in this scenario, $m_{\chi_4^0} \sim \chi_2^\pm \le  520 \, {\rm GeV}$, whereas the ATLAS Collaboration
 could only exclude points with Wino mass smaller than 380 GeV \cite{ATLAS:2017uun}. The CMS Collaboration
has better results, but only excludes the points with Wino mass smaller than 450 GeV \cite{CMS:2017fdz}. One can also see \cite{Baer:2014kya,Han:2013usa} for probing light higgsino using monojet searches.
This implies 
that in our case, even though the Higgsino LSP solutions are natural solutions but are out of reach of the present 
LHC searches. And we have confirmed it from numerical calculations for LHC SUSY searches as well.
It is therefore, we will not consider them for further analyses. 

We want to comment on the light stau-neutralino coannihilation solutions. We find that if we insisting 
on $\Delta_{EW}\lesssim$ 50, the light stau-neutralino coannihilation scenario is knocked out though 
it can be achieved if we relax $\Delta_{EW}$ up to 100. This is why we will consider only $Z$-pole and 
Higgs-pole solutions for collider studies. 

We have collected five represented benchmark points (BMP) in Table 1. BMP1 and BMP2 are
the examples of $Z$-pole solutions with correct relic density and small relic density respectively.
In these points, gluinos and the first two generations of squarks are heavier than 2 TeV, while
 top and bottom squarks $\tilde t_{1,2}$ and $\tilde b_{1,2}$ have masses from 1 TeV to 2 TeV. 
The first two generations of sleptons have masses are in the range of [800, 1000] GeV
while $\tilde \tau_{1,2}$ are in [720, 950] GeV mass range. For BMP1 and BMP2, $\Delta_{EW}$ is 29 and 24 while $m_{\tilde \chi_{1}^{\pm}}$ is 333 GeV and 260 GeV, respectively. BMP3 and BMP4 represent Higgs-pole
solutions with correct relic density and small relic density, respectively. The colored sparticles have
masses in same range as for BMP1 and BMP2. The first two generation sleptons are in the mass range
[520, 790] GeV while $\tilde \tau_{1,2}$ are in [160, 790] GeV. BMP5 show a higgsino LSP solution. This point also has similar spectrum as BMP1 with relatively heavy sleptons and winos are heavier than 550 GeV. Since these points have very small relic density ($\Omega h^2\sim 0.002$), we rescale the direct detection rate as $\xi \sigma^{SI}(\chi,p)$.  
Moreover, we can see that BMP2, BMP3 and BMP4 have $\Delta a_{\mu}$ within 3$\sigma$~\cite{Bennett:2006fi}.


\section{LHC searches}\label{lhc_searches}

In this Section, we examine the constraints from the direct searches for the SUSY particles at the LHC on samples with relic density consistent with or smaller than $5\sigma$ WMAP9 bounds and also satisfy the current LUX limits on direct detection of LSP neutralino.
In order to realize small fine-tuning and satisfy experimental observations simultaneously, only electroweakinos (neutralinos and charginos) and stau are light and could be explored at the current LHC.
Therefore, we should consider the following electroweak Drell-Yan production processes:
\begin{equation}
	pp \to \tilde{\chi}_i^0 \tilde{\chi}_j^0, \hspace{4mm} pp \to \tilde{\chi}_l^\pm \tilde{\chi}_m^\pm, \hspace{4mm} pp \to \tilde{\chi}_i^0 \tilde{\chi}_l^\pm,
	\hspace{4mm} (i,j=2,3,4~\mathrm{and}~l,m=1,2).
\end{equation}

In general, the produced neutralinos could decay through on-shell or off-shell  $Z^{(\ast)}$ or $h^{(\ast)}$:
\begin{equation}
	\tilde{\chi}_i^0 \to	Z^{(\ast)} \chia, \hspace{6mm} \tilde{\chi}_i^0 \to h^{(\ast)} \chia,
\end{equation}
where the charginos could only decay through $W^{(\ast)}$,
\begin{equation}
	\tilde{\chi}_l^\pm \to W^{\pm {(\ast)}} \chia.
\end{equation}

When $\staua$ is light, such as BMP 3 and BMP 4, new decay modes of $\tilde{\chi}_i^0$ and $\tilde{\chi}_l^\pm$ are possible,
\begin{equation}
	\tilde{\chi}_i^0 \to \staua \tau, \hspace{4mm} \mathrm{and}, \hspace{4mm}
	\tilde{\chi}_l^\pm \to \staua v_{\tau},
\end{equation}
and $\staua$ decay into $\chia$ with branching ratio approximates to $100\%$.

\begin{figure}[!htb]
	\centering
	\subfigure[$\chib$ decay branching ratio, Z pole.\label{fig:br_a}]{
	\includegraphics[width=0.45\textwidth]{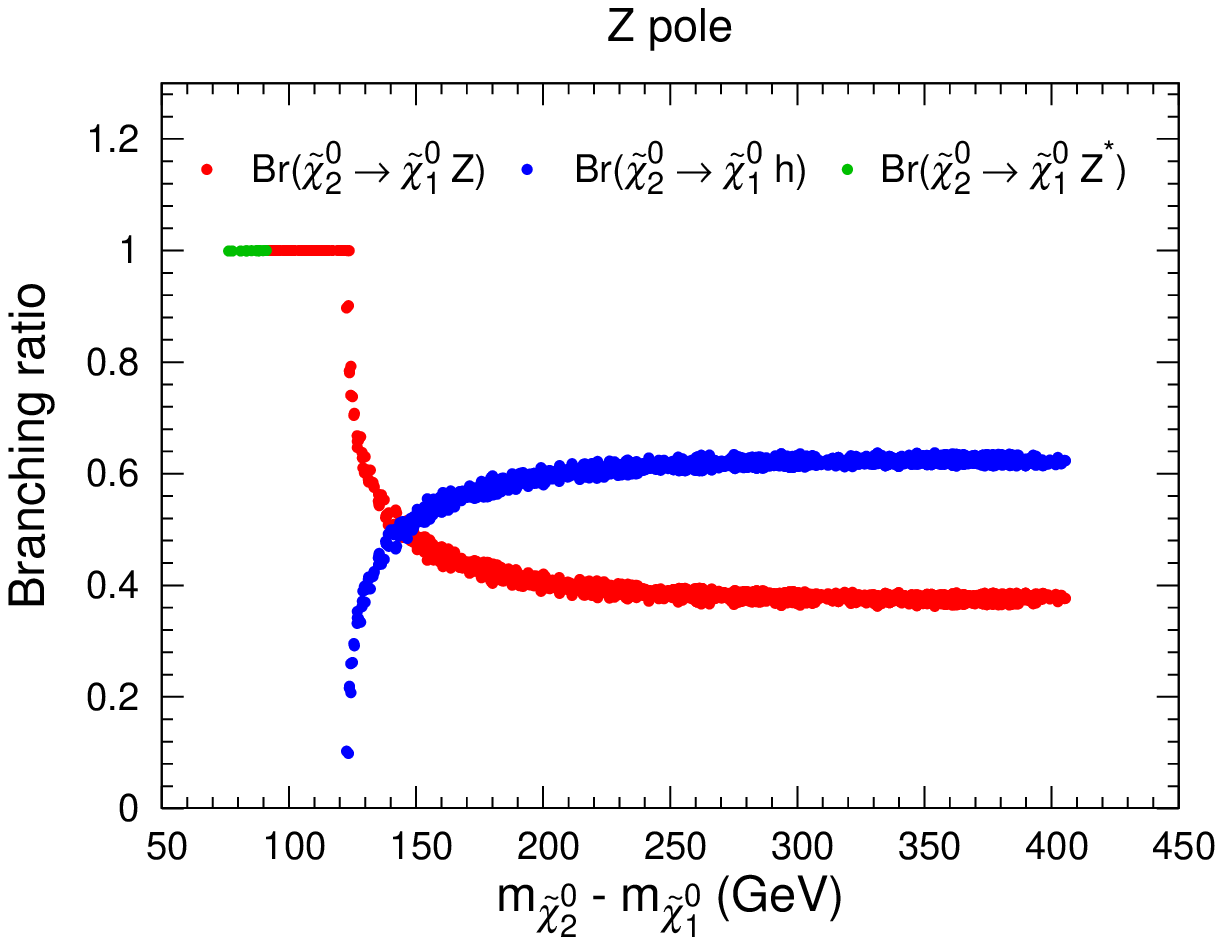}}
	\subfigure[$\chiapm$ decay branching ratio, Z pole.\label{fig:br_b}]{
	\includegraphics[width=0.45\textwidth]{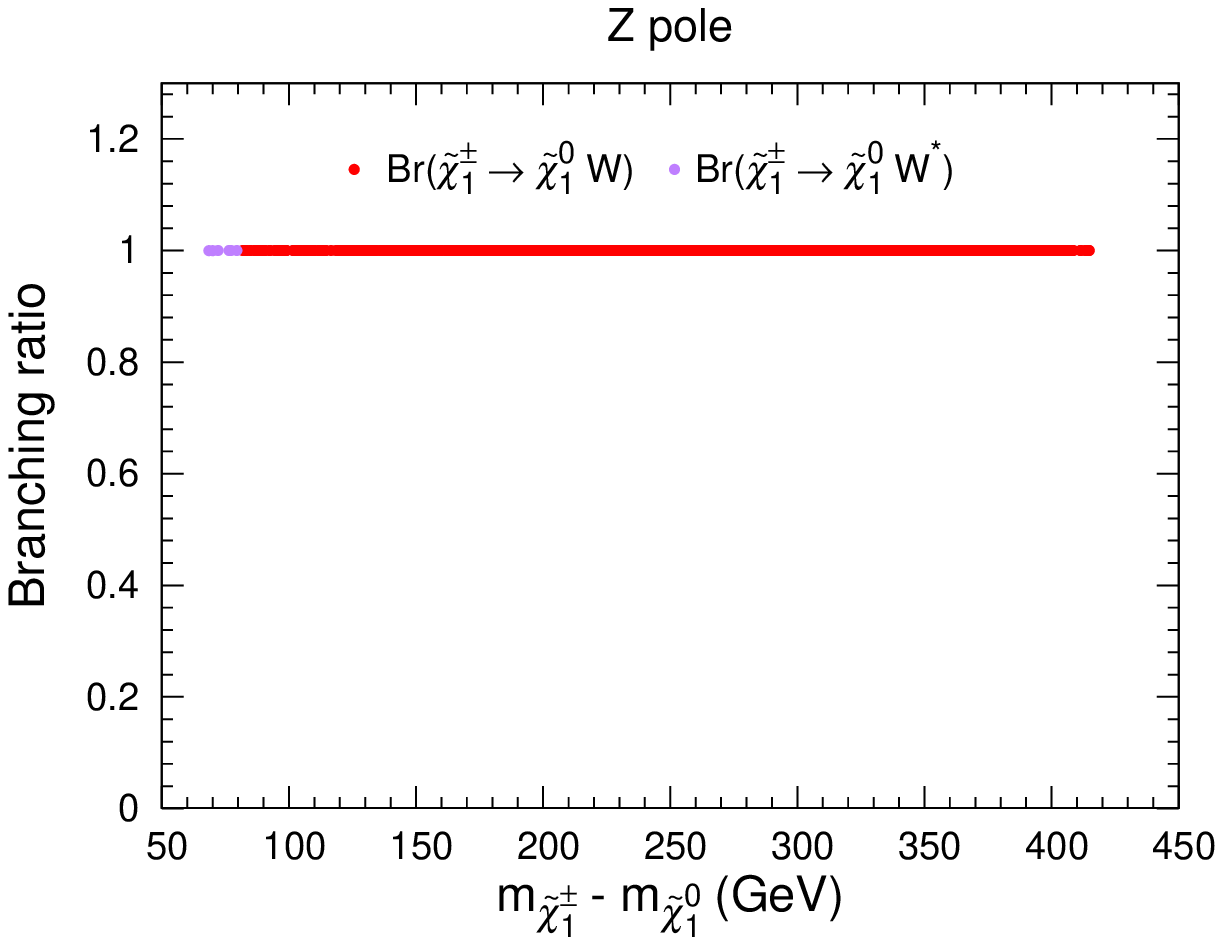}}
	\subfigure[$\chib$ decay branching ratio, H pole.\label{fig:br_c}]{
	\includegraphics[width=0.45\textwidth]{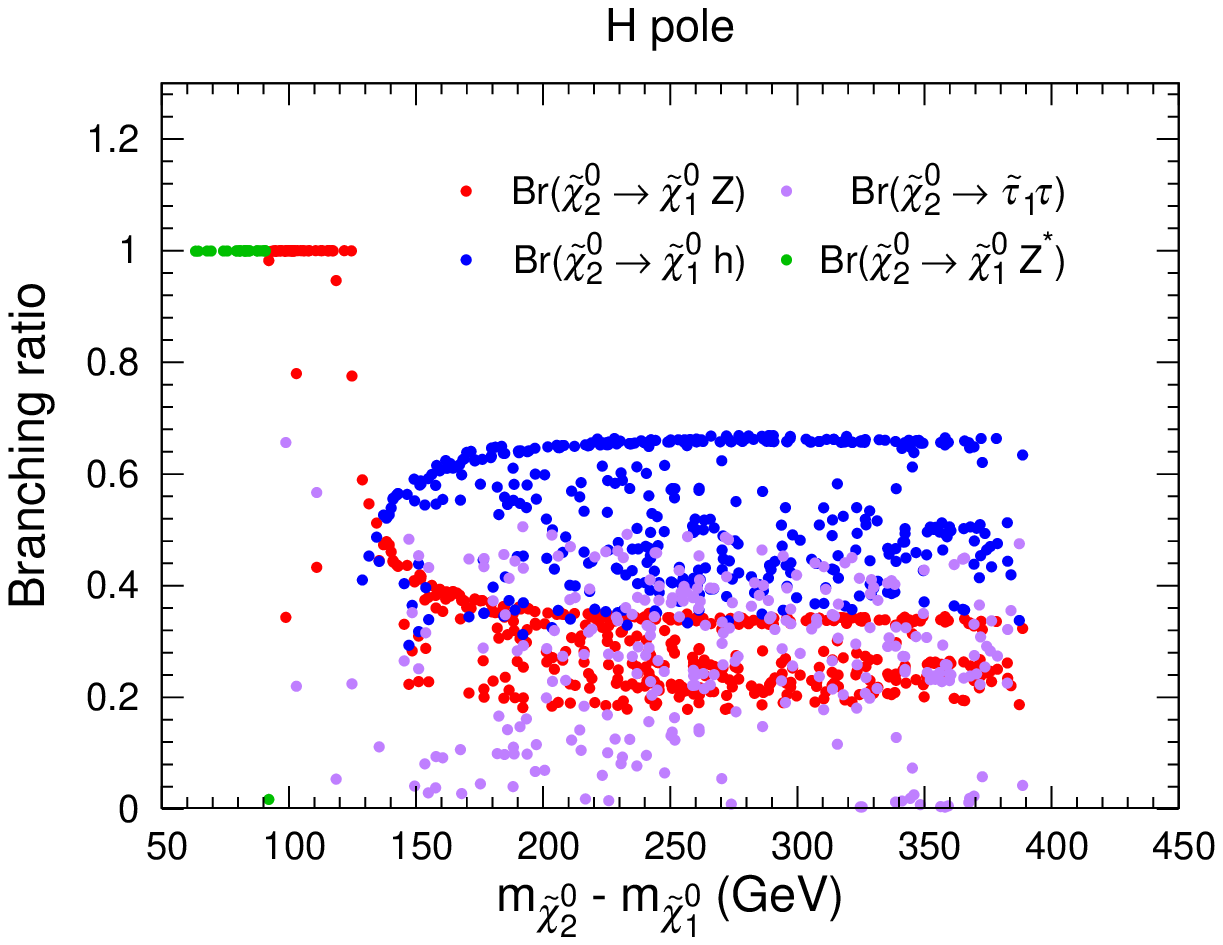}}
	\subfigure[$\chiapm$ decay branching ratio, H pole.\label{fig:br_d}]{
	\includegraphics[width=0.45\textwidth]{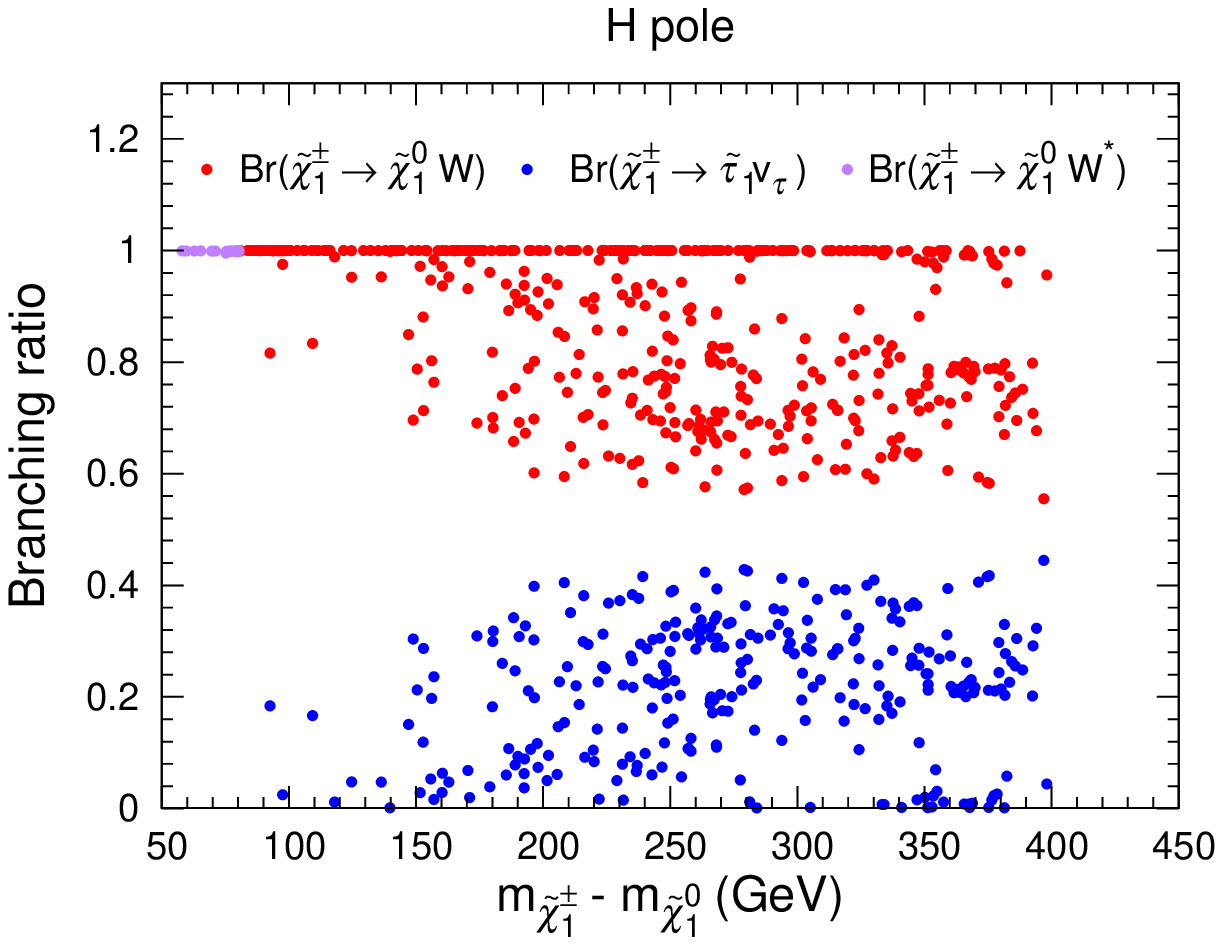}}
	\caption{Decay branching ratios of $\chib$ (a and b) and $\chiapm$ (c and d) for samples with relic density consistent with or smaller than $5\sigma$ WMAP9 bounds.
	}
	\label{fig:branch}
\end{figure}

The relevance of different decay modes depends on mass spectrums and will significantly influence collider searches for these particles.
In Fig.~\ref{fig:branch} we show branching ratios of $\chib$ (Figs.~\ref{fig:br_a} and \ref{fig:br_c}) and $\chiapm$ (Figs.~\ref{fig:br_b} and \ref{fig:br_d}) for samples considered in this work.
The dominant decay channel of $\chib$ for samples of $Z$ pole is $\chib \to \chia Z^{(\ast)}$ when the mass difference $m_{\chib}-m_{\chia}$ is small.
Once the decay into Higgs boson is kinematically possible, branching ratio to $\chia h$ increase with increasing of $m_{\chib}-m_{\chia}$ and become the dominant channel when $m_{\chib}-m_{\chia} \gtrsim 140~\GeV$.
The decay channels of $\chiapm$ is always $\chiapm \to \chia W^{(\ast)}$.
For samples of Higgs pole, situations are more complex due to light $\staua$, as can be seen from BMP3 and BMP4.
The decay of $\chib$ to $\tilde{\tau}_1\tau$ would be significant or even dominant~\ref{fig:br_c}. 


ATLAS~\cite{ATLAS:2016uwq, Aaboud:2017nhr} and CMS~\cite{CMS:2017fdz} have performed electroweakinos searches for wino $\chib/\chiapm$ with particular decay models.
We use the powerful package \texttt{CheckMate}~\cite{Drees:2013wra,Kim:2015wza,Dercks:2016npn} (where \texttt{PYTHIA~8}~\cite{Sjostrand:2014zea} and a tuned version of \texttt{DELPHES~3}~\cite{deFavereau:2013fsa} have been used internally) 
to implement LHC constraints. 
NLO production rates are obtained by rescaling LO rates with K-factors calculated by \texttt{Prospino~2}~\cite{Beenakker:1999xh}, which yield about 1.2 for higgsino pair production.

As for electroweakinos searches, currently \texttt{CheckMate} has only employed
the ATLAS analyses with $13.3~\mathrm{fb}^{-1}$ data~\cite{ATLAS:2016uwq}.
So in order to fully take into account the current constraints, we also recast the latest ATLAS~\cite{Aaboud:2017nhr} and CMS~\cite{CMS:2017fdz} analyses based on a Monte Carlo simulation.
In the simulation, \texttt{MadGraph~5}~\cite{Alwall:2014hca} is adopted to generate background and signal samples, and \texttt{PYTHIA~6}~\cite{Sjostrand:2006za} is employed to handle the parton shower, particle decay, and hadronization processes.
We use MLM scheme to deal with the matching between matrix element and parton shower calculations, and use \texttt{Delphes~3}~\cite{deFavereau:2013fsa} to carry out a fast detection simulation with the CMS setup. 
Jets are reconstructed using the $\mathrm{anti}-k_T$ algorithm~\cite{Cacciari:2008gp} with a distance parameter $ \Delta R=0.4 $.


Generally, heavy electroweakinos productions with successive decay will lead to multi-leptons signal, among them $3l+\missET$ and $2l+\missET$ give the best sensitivity at the LHC searches.
In the case of $3l + \missET$ search channel, major SM backgrounds are  $ZZ$ and $WZ$ productions.
Two leptons from $Z$ decay are required to form same-flavor-opposite-sign (SFOS) pair.
Two useful kinematic variables to discriminate signals from backgrounds are $\mT$ and $\missET$, where $\mT$ is the transverse mass defined as $\mT = \sqrt{ 2 (p_\mathrm{T}^l \missET - \mathbf{p}_\mathrm{T}^{l} \cdot \mathbf{p}_\mathrm{T}^\mathrm{miss})}$ with ${p}_\mathrm{T}^\mathrm{miss}$ is the missing transverse momentum vector and the lepton $l$ is the one not forming the SFOS lepton pair.

\begin{figure}[!htb]
	\centering
	\subfigure[~$\mT$ distributions~\label{fig:3l_bmp:mT}]{
	\includegraphics[width=0.45\textwidth]{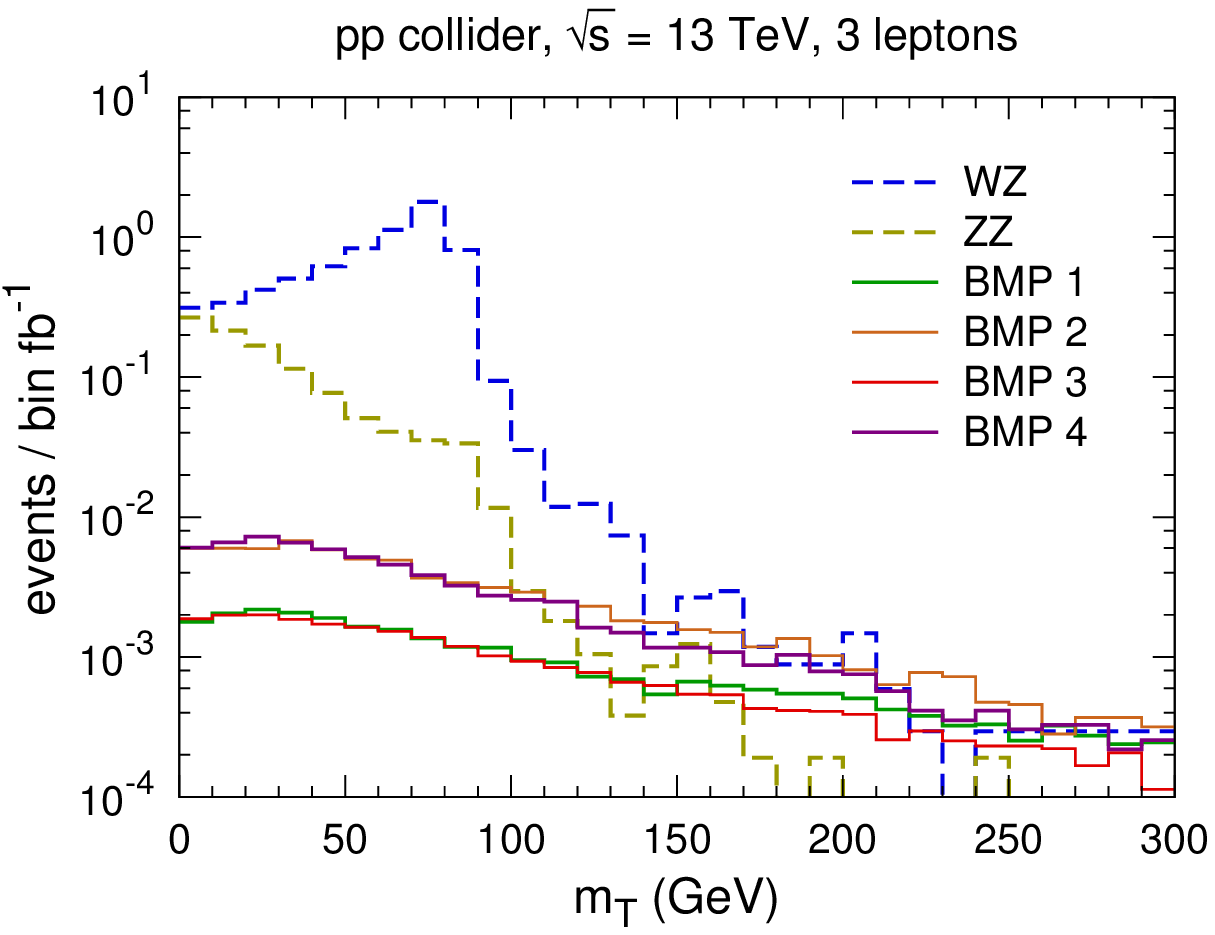}}
	\subfigure[~$\missET$ distributions~\label{fig:3l_bmp:missET}]{
	\includegraphics[width=0.45\textwidth]{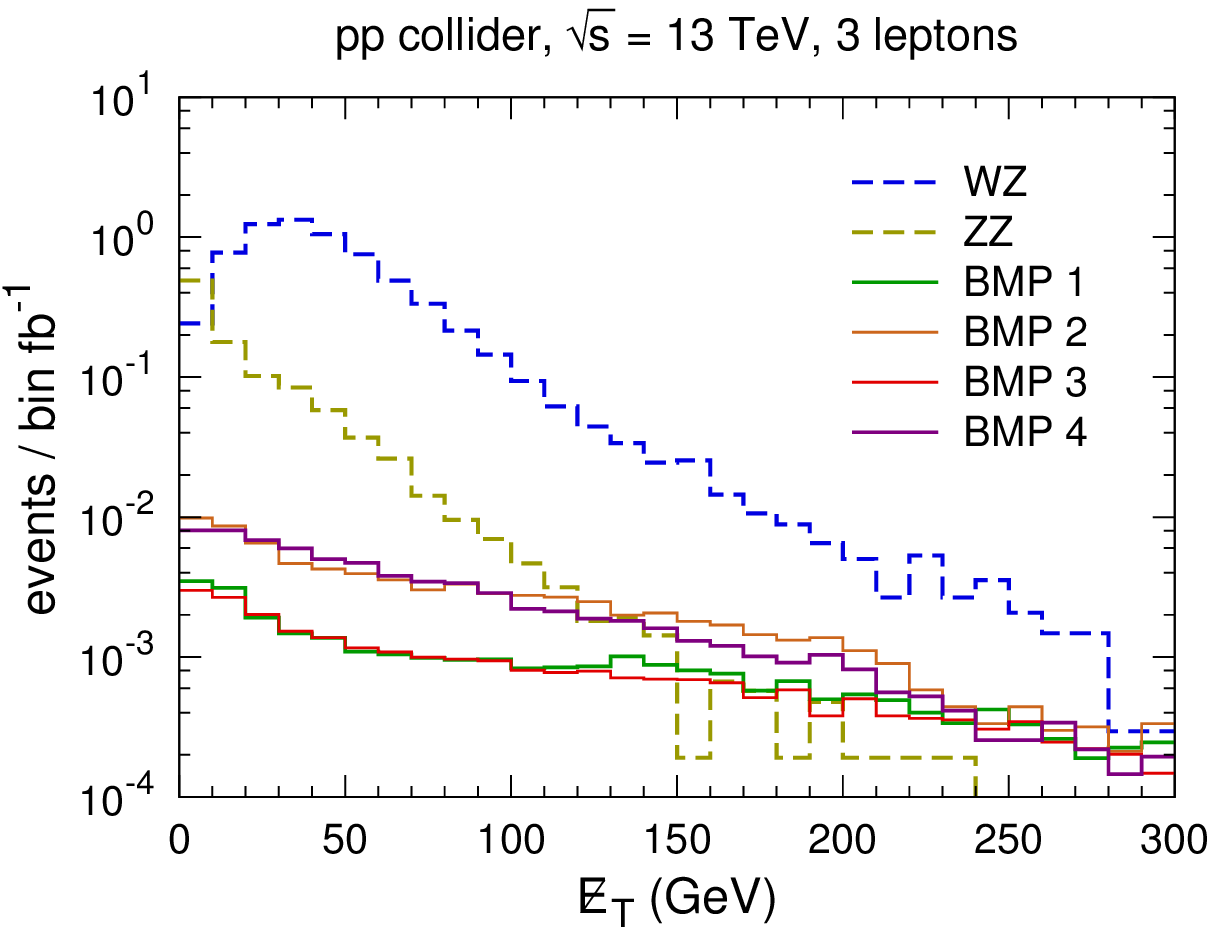}}
	\caption{$\mT$ (a) and $\missET$ (b) distributions for backgrounds and signal benchmark points in the $2l + \missET$ channel at the 13~TeV LHC.
	}
	\label{fig:3l_bmp}
\end{figure}

In Fig.~\ref{fig:3l_bmp} we present the $\mT$ and $\missET$ distributions of backgrounds and signals.
In the case of $ZZ$ background, the $3l$ final state mainly comes from the decay of both $Z$ boson into $l^+l^-$ pairs with one lepton do not  be successfully reconstructed. 
As there is no neutrino contributing $\missET$, its $\missET$ distribution is softer than others, and so is its $\mT$ distribution.
For the $WZ$ background, the $\mT$ variable is bounded by the $W$ boson mass, leading to an obvious endpoint near $m_W$.
All $\mT$ and $\missET$ for signals are harder and are easy to be distinguished from backgrounds.


In the case of $2l+\missET$ search channel, dominant backgrounds are $WZ$, $WW$, $ZZ$, and $t\bar{t}$ production.
Still, these two leptons from $Z$ decay are required to form SFOS pairs, whose invariable is a useful variable to distinguish signals from SM backgrounds.
Another useful variable $\mTT$ is defined as
\begin{equation}
  \mTT = \underset{\mathbf{p}_\mathrm{T}^1 + \mathbf{p}_\mathrm{T}^2 =  \mathbf{p}_\mathrm{T}^\mathrm{miss}}{\mathrm{min}}
  \{\mathrm{max}[m_\mathrm{T}(\mathbf{p}_\mathrm{T}^a,\mathbf{p}_\mathrm{T}^1), m_\mathrm{T}(\mathbf{p}_\mathrm{T}^b,\mathbf{p}_\mathrm{T}^2) ] \} ,
  \label{eq:mT2}
\end{equation}
where $m_\mathrm{T}(\mathbf{p}_\mathrm{T}^i,\mathbf{p}_\mathrm{T}^j)=\sqrt{2(p_\mathrm{T}^i p_\mathrm{T}^j
-\mathbf{p}_\mathrm{T}^i\cdot\mathbf{p}_\mathrm{T}^j)}$, and $\mathbf{p}_\mathrm{T}^a$ and $ \mathbf{p}_\mathrm{T}^b$ are the transverse momenta of two visible particles in the decay chain (two leptons in our case).
$\mathbf{p}_\mathrm{T}^1$ and $\mathbf{p}_\mathrm{T}^2$ are a partition of the missing transverse momentum $\mathbf{p}_\mathrm{T}^\mathrm{miss}$.
By definition, $\mTT$ is the minimum of the larger $m_\mathrm{T}$ over all partitions, its distribution for two identical chains has an upper endpoint, which is determined by the mass difference between the parent particle and its invisible child.

\begin{figure}[!htb]
	\centering
	\subfigure[~$\mathrm{m}_{ll}$ distributions~\label{fig:2l_bmp_mll}]{
	\includegraphics[width=0.45\textwidth]{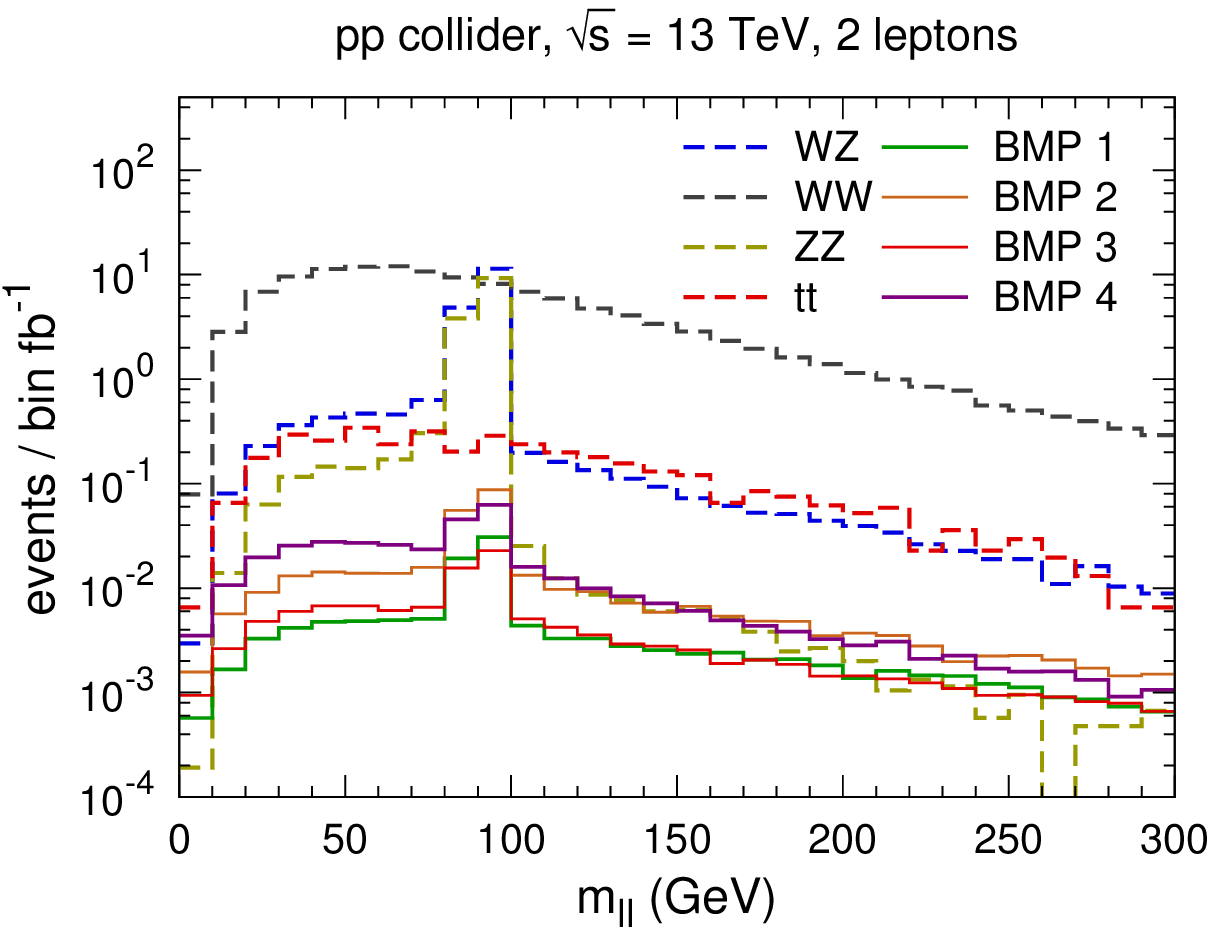}}
	\subfigure[~$\mTT$ distributions~\label{fig:2l_bmp_mT2}]{
	\includegraphics[width=0.45\textwidth]{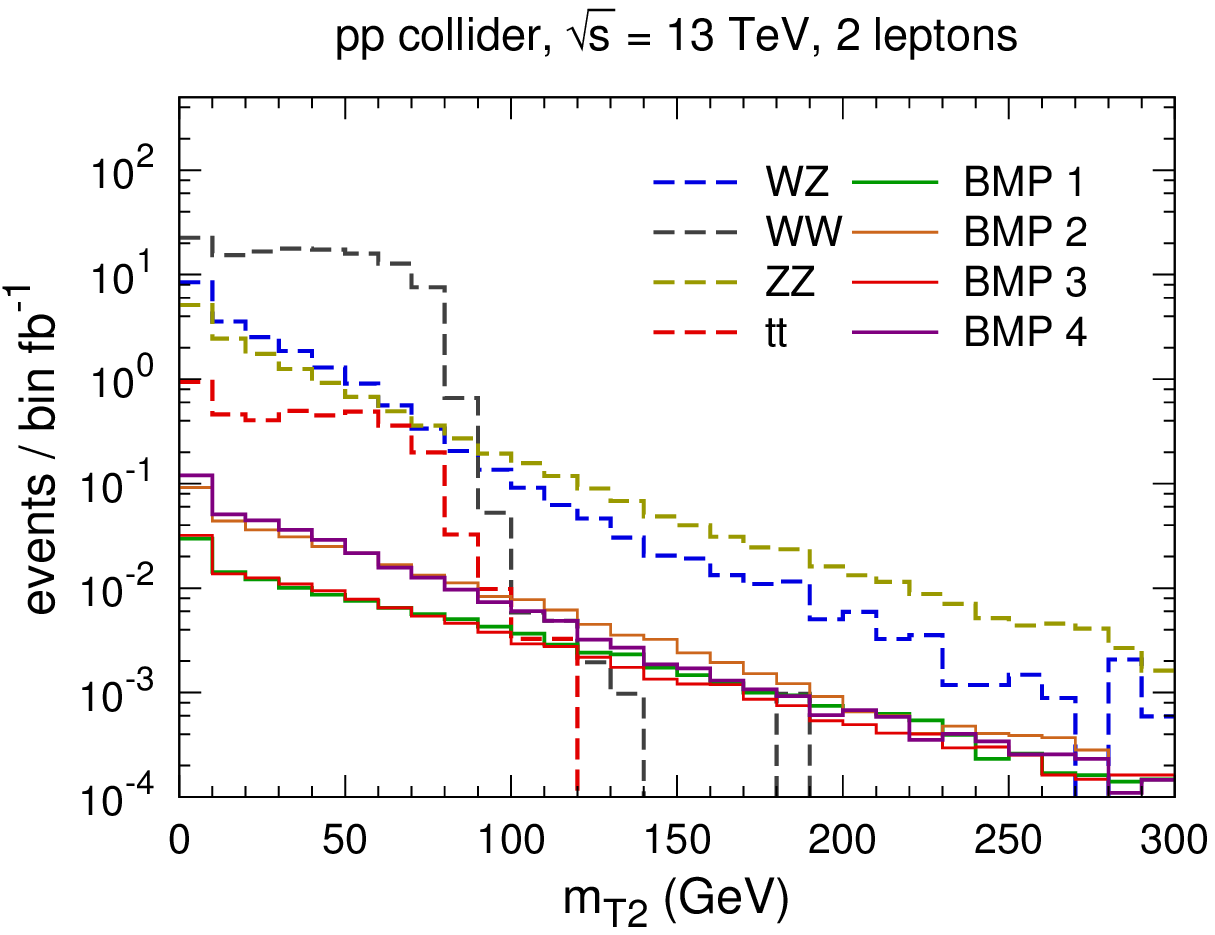}}
	\caption{$\mathrm{m}_{ll}$ (a) and $\mTT$ (b) distributions for backgrounds and signal benchmark points in the $2l + \missET$ channel at the 13~TeV LHC.
	}
	\label{fig:2l_bmp}
\end{figure}

In Fig.~\ref{fig:2l_bmp} we demonstrate the $\mathrm{m}_{ll}$ and $\mTT$ distributions of backgrounds and signals.
For the $WZ$ and $ZZ$ backgrounds and signals, lepton pairs from  $Z$ boson decay result in peaks around $m_Z$ in the $\mathrm{m}_{ll}$ distributions, as shown in Fig.~\ref{fig:2l_bmp_mll}.
Whereas these two leptons for $WW$ and $t\bar{t}$ backgrounds origin from two particles and do not have obvious feature.
Fortunately, the $\mTT$ distributions for the $WW$ and $t\bar{t}$ backgrounds are essentially bounds by $m_W$.

\begin{figure}[!htbp]
	\centering
	\subfigure[$m_{\chia}$-$m_{\chib}$ plane~\label{fig:mn1_mn2}]{
	\includegraphics[width=0.45\textwidth]{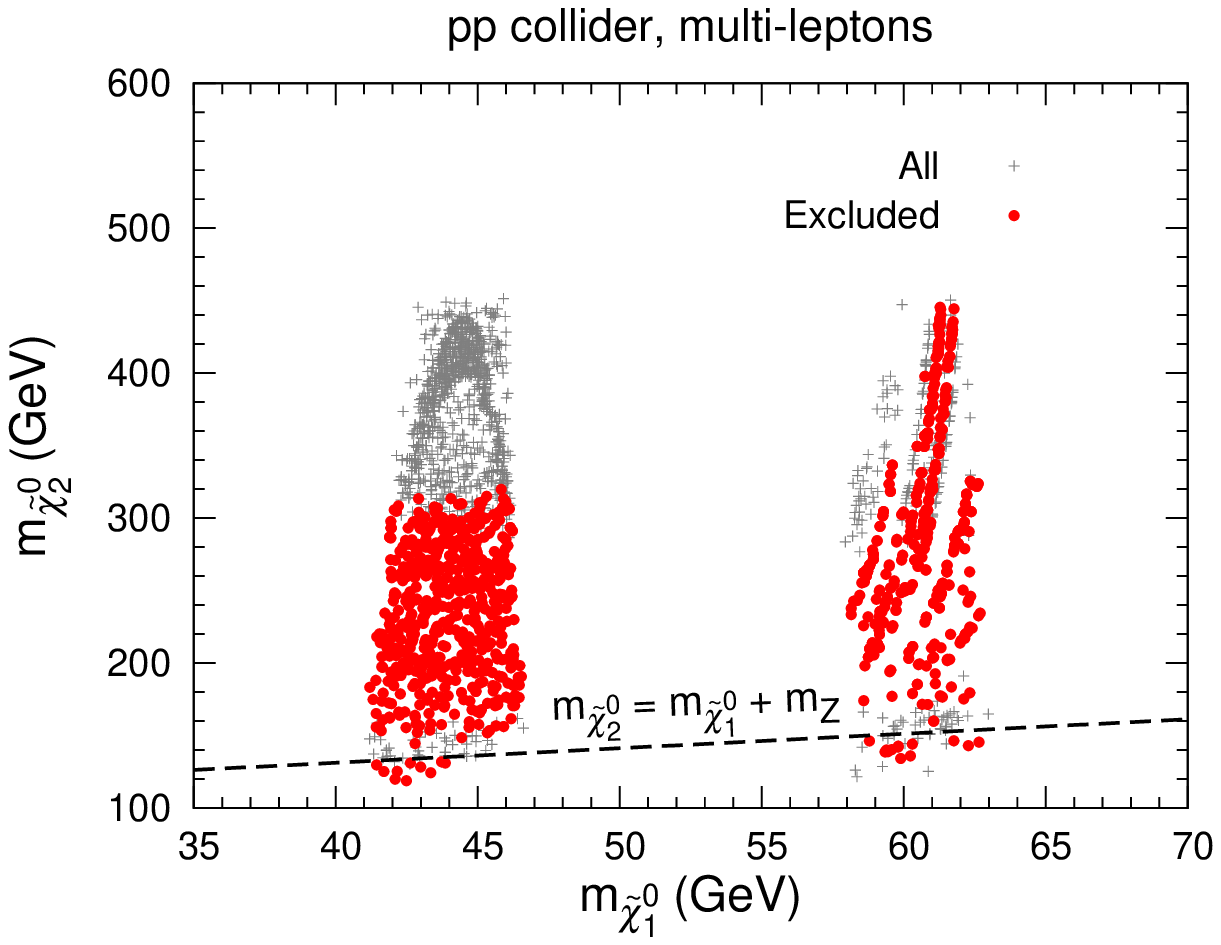}}
	\subfigure[$m_{\chia}$-$\Delta_{\mathrm{EW}}$ plane~\label{fig:deltaEW}]{
	\includegraphics[width=0.45\textwidth]{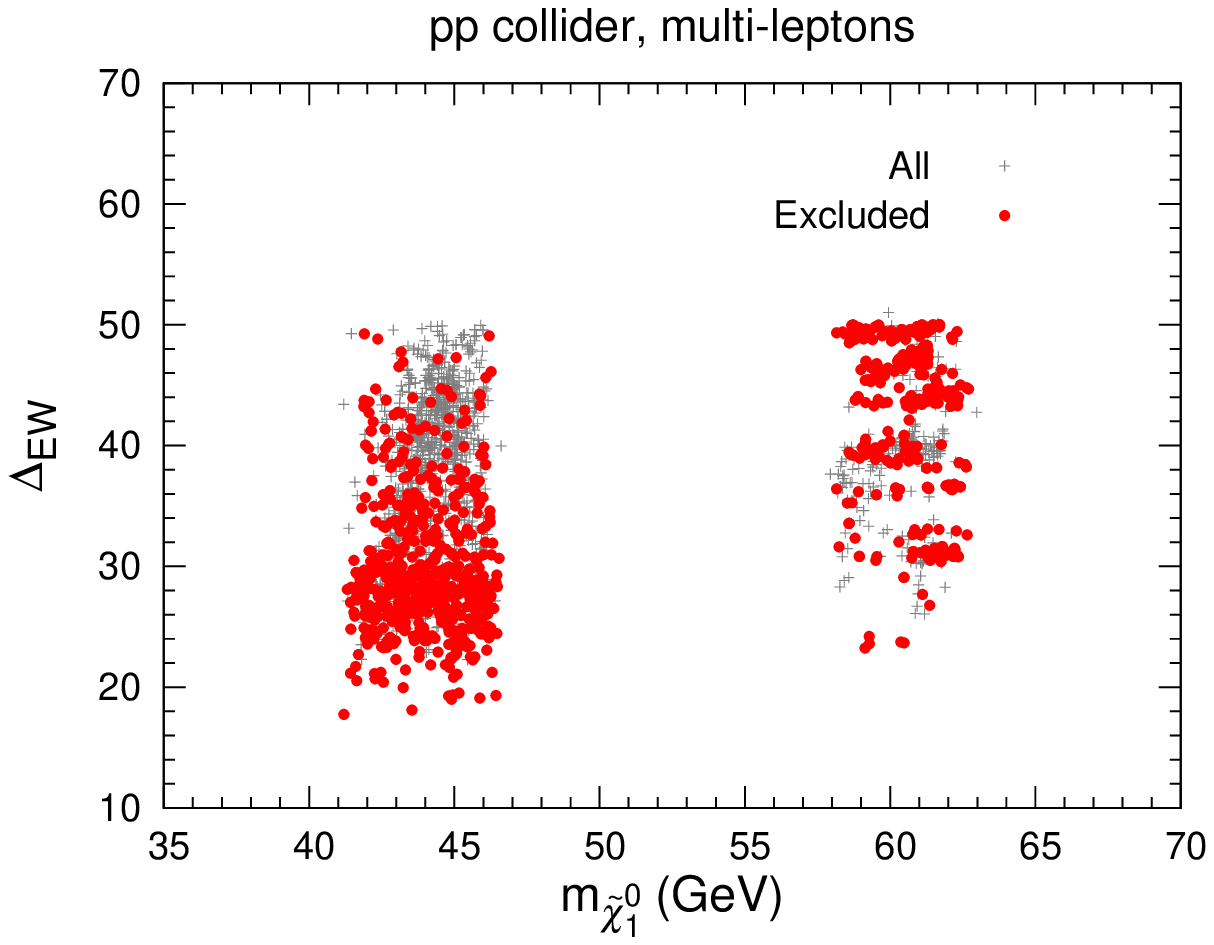}}
	\caption{$95\%$ C.L. exclusion results of the LHC electroweakinos searches in the $m_{\chia}$-$m_{\chib}$ plane (a) and $m_{\chia}$-$\Delta_{\mathrm{EW}}$ plane (b).
	}
	\label{fig:result}
\end{figure}

For the analyses of Ref.~\cite{ATLAS:2016uwq}, we use \texttt{CheckMATE} to calculate corresponding significance.
And for the  analyses of Refs.~\cite{CMS:2017fdz} and Ref.~\cite{Aaboud:2017nhr}, we apply the same cuts 
in various signal regions, and compare the obtained cross sections to $95\%$ limits tabulated in these literatures.
In Fig.~\ref{fig:result}, we present the $95\%$ C.L. exclusion results of the LHC electroweakino searches in the $m_{\chia}$-$m_{\chib}$ plane (Fig.~\ref{fig:mn1_mn2}) and $m_{\chia}$-$\Delta_{\mathrm{EW}}$ plane (Fig.~\ref{fig:deltaEW}). 
$3l + \missET$ searches require exactly three hard leptons.
As a result, samples close to $Z$ threshold, i.e., $m_{\chib} \sim m_{\chia} + m_Z$, are hard to explore.
Points below this threshold would be excluded by $2l+\missET$ searches due to small $m_{\chiapm}$ and large production cross sections.

Roughly, $3l+\missET$ and $2l + \missET$ searches at the current LHC could exclude higgsino dominant $\chib/\chiapm$ with mass up to $300~\GeV$.
This result is consistent with our previous prediction~\cite{Xiang:2016ndq} whereas seems to somewhat weaker than 
that given by the ATLAS~\cite{ATLAS:2016uwq} and CMS Collaborations~\cite{CMS:2017fdz}. 
The main reason is that in their searches, pure wino $\chib/\chiapm$ is assumed, which have larger production cross sections. 
Besides, they assume $\chiapm$ and $\chib$ decay via $W^{\ast}$ and $Z^{\ast}$ bosons with a branching fraction of $100\%$, whereas decay branching ratios in our samples highly depend on mass spectra.

Another and even stricter constraint on samples of Higgs pole come from the searches for electroweakinos 
with tau final states.
These searches have been performed by both CMS~\cite{CMS:2017fdz} and ATLAS~\cite{Aaboud:2017nhr} Collaborations, 
by latter it was shown that  when $\chib/\chiapm$ decay into $\chia$ via an intermediate on-shell stau or tau sneutrino, $\chib/\chiapm$ with mass up to $760~\GeV$ are excluded for a massless $\chia$.
In our case samples with $\chib$ mass up to $460~\GeV$ for Higgs pole could still be excluded, as shown in Fig.~\ref{fig:mn1_mn2}.

Finally, we project exclusion results into $m_{\chia}$-$\Delta_{\mathrm{EW}}$ plane (Fig.~\ref{fig:deltaEW}).
Samples of $Z$ pole with small $\Delta_{\mathrm{EW}}$ are easy to be explored, whereas these with large $\Delta_{\mathrm{EW}}$ are hard to be excluded due to large $\mu$, which in turn indicate large $m_{\chib}$ and small production cross sections.
In the case of Higgs pole, many samples with $\Delta_{\mathrm{EW}}$ up to 50 could by excluded by electroweakinos searches with tau final states.
For both $Z$ pole and Higgs pole, samples with $\Delta_{\mathrm{EW}}$ approximate to 20 could still survive, indicating naturalness of this SUSY framework.

\section{Discussions and Conclusion}\label{dc}

We have studied natural supersymmetry in the GmSUGRA, and found that after demanding $\Delta_{EW}\lesssim$ 50,    
only the parameter space related to $Z$-pole and Higgs-pole solutions are left. We performed the focused scans 
for such parameter space and showed that it satisfies various phenomenological constraints and is compatible 
with the current direct detection bound on neutralino DM reported by the LUX experiment. Such parameter space 
also has solutions with the correct DM relic density besides the solutions with relic density smaller 
or larger than 5$\sigma$ WMAP9 bounds. We also performed the collider study of such solutions by implementing 
and comparing with relevant studies done by the ATLAS and CMS Collaborations. We showed that the points with
the higgsino dominant $\chib/\chiapm$ mass up to $300~\GeV$ are excluded for $Z$ pole scenario 
while for Higgs-pole scenario, points with $\chib$ mass up to $460~\GeV$ are excluded. Next, 
we displayed that both for the Z-pole and Higgs-pole scenarios, the points having $\Delta_{EW}\sim$ 20 still survive. 
Moreover, we present five benchmark points as examples of our present scans. In these benchmark points,
 gluino and the first two generations of squarks are heavier than 2 TeV, top squarks $\tilde t_{1,2}$ are 
in the mass range $[1,~2]$ TeV, while sleptons are lighter than 1 TeV. We also discuss that stau-neutralino coannihilation scenario is not compatible with our demand of $\Delta_{EW}\lesssim$ 50. On the other hand higgsino LSP solutions which are natural solutions but are out of reach of present LHC searches. Some part of the parameter space can explain the anomaly of muon $(g-2)_{\mu}$ within 3$\sigma$ as well.

\section*{Acknowledgements}

(XJB) and (PF) were supported by National Natural Science Foundation of China under Grants No.11475189, 11475191, and the National Key Program for Research and Development (No. 2016YFA0400200).
(TL) was supported in part 
by the Projects 11475238 and 11647601 supported
by National Natural Science Foundation of China, 
and by Key Research Program of Frontier Science, CAS and the CAS-TWAS President’s Fellowship
Programme (WA). The numerical results described in this paper have been obtained via the HPC Cluster of ITP-CAS.
(SR) likes to thank (TL) for warm hospitality and Institute of Theoretical Physics, 
CAS, Beijing, China for providing conducive atmosphere for research 
where part of this work has carried out.



\begin{thebibliography}{99}
\bibitem{gaugeunification} S.~Dimopoulos, S.~Raby and F.~Wilczek,
  Phys.\ Rev.\ D {\bf 24} (1981) 1681;
U.~Amaldi, W.~de Boer and H.~Furstenau,
  Phys.\ Lett.\ B {\bf 260}, 447 (1991);
J.~R.~Ellis, S.~Kelley and D.~V.~Nanopoulos,
  Phys.\ Lett.\ B {\bf 260} (1991) 131;
P.~Langacker and M.~X.~Luo,
  Phys.\ Rev.\ D {\bf 44} (1991) 817.
\bibitem{ghp} E.~Witten, 
  Nucl.\ Phys.\ B {\bf 188}, 513 (1981); R.~K.~Kaul, 
Phys.\ Lett.\ B {\bf 109}, 19 (1982).

\bibitem{mhiggs} H.~E.~Haber and R.~Hempfling,
  Phys.\ Rev.\ Lett.\  {\bf 66} (1991) 1815;
J.~R.~Ellis, G.~Ridolfi and F.~Zwirner,
  Phys.\ Lett.\ B {\bf 257} (1991) 83;
Y.~Okada, M.~Yamaguchi and T.~Yanagida,
  Prog.\ Theor.\ Phys.\  {\bf 85} (1991) 1;
For a review, see {\it e.g.}  M.~S.~Carena and H.~E.~Haber,
  Prog.\ Part.\ Nucl.\ Phys.\  {\bf 50} (2003) 63
  [hep-ph/0208209].



\bibitem{Aad:2012tfa}
  G.~Aad {\it et al.}  [ATLAS Collaboration],
  Phys.\ Lett.\ B {\bf 716}, 1 (2012)
  [arXiv:1207.7214 [hep-ex]].

\bibitem{CMS} 
  S.~Chatrchyan {\it et al.}  [CMS Collaboration],
  Phys.\ Lett.\ B {\bf 716}, 30 (2012)
  [arXiv:1207.7235 [hep-ex]].

\bibitem{neutralinodarkmatter}
H.~Goldberg, 
  Phys.\ Rev.\ Lett.\  {\bf 50}, 1419 (1983);
J.~Ellis, J.~Hagelin, D.V.~Nanopoulos, K.~Olive, and M.~Srednicki,
  Nucl.\ Phys.\ B {\bf 238}, 453 (1984).

\bibitem{darkmatterreviews} For reviews, see
G.~Jungman, M.~Kamionkowski and K.~Griest   
  Phys.\ Rept.\  {\bf 267}, 195 (1996)
  [hep-ph/9506380];
K.A.~Olive,
  ``TASI Lectures on Dark matter,''
  [astro-ph/0301505];
J.L.~Feng,
  ``Supersymmetry and cosmology,'' 
  [hep-ph/0405215];
M.~Drees,
  ``Neutralino dark matter in 2005,''
  [hep-ph/0509105].
J.L.~Feng,
  Ann.\ Rev.\ Astron.\ Astrophys.\  {\bf 48}, 495 (2010)
  [arXiv:1003.0904 [astro-ph.CO]].


\bibitem{lhc-strong} ATLAS collaboration, ATLAS-CONF-2017-022; CMS
  Collaboration, CMS-SUS-16-036.



\bibitem{Drees:2015aeo} 
  M.~Drees and J.~S.~Kim,
  Phys.\ Rev.\ D {\bf 93}, no. 9, 095005 (2016)
  [arXiv:1511.04461 [hep-ph]].

\bibitem{Ding:2015epa} 
  R.~Ding, T.~Li, F.~Staub and B.~Zhu,
  Phys.\ Rev.\ D {\bf 93}, no. 9, 095028 (2016)
  [arXiv:1510.01328 [hep-ph]].

\bibitem{Baer:2015rja} 
  H.~Baer, V.~Barger and M.~Savoy,
  Phys.\ Rev.\ D {\bf 93}, no. 3, 035016 (2016)
  [arXiv:1509.02929 [hep-ph]].

\bibitem{Batell:2015fma} 
  B.~Batell, G.~F.~Giudice and M.~McCullough,
  JHEP {\bf 1512}, 162 (2015)
  [arXiv:1509.00834 [hep-ph]].

\bibitem{AbdusSalam:2015uba} 
  S.~S.~AbdusSalam and L.~Velasco-Sevilla,
  Phys.\ Rev.\ D {\bf 94}, no. 3, 035026 (2016)
  [arXiv:1506.02499 [hep-ph]].

\bibitem{Barducci:2015ffa} 
  D.~Barducci, A.~Belyaev, A.~K.~M.~Bharucha, W.~Porod and V.~Sanz,
  JHEP {\bf 1507}, 066 (2015)
  [arXiv:1504.02472 [hep-ph]].
\bibitem{Cohen:2015ala} 
  T.~Cohen, J.~Kearney and M.~Luty,
  Phys.\ Rev.\ D {\bf 91}, 075004 (2015)
  [arXiv:1501.01962 [hep-ph]].

\bibitem{Fan:2014axa} 
  J.~Fan, M.~Reece and L.~T.~Wang,
  JHEP {\bf 1508}, 152 (2015)
  [arXiv:1412.3107 [hep-ph]].

\bibitem{Leggett:2014hha} 
  T.~Leggett, T.~Li, J.~A.~Maxin, D.~V.~Nanopoulos and J.~W.~Walker,
  arXiv:1403.3099 [hep-ph];
  Phys.\ Lett.\ B {\bf 740}, 66 (2015)
  [arXiv:1408.4459 [hep-ph]].

\bibitem{Dimopoulos:2014aua} 
  S.~Dimopoulos, K.~Howe and J.~March-Russell,
  Phys.\ Rev.\ Lett.\  {\bf 113}, 111802 (2014)
  [arXiv:1404.7554 [hep-ph]].
\bibitem{Gogoladze:2013wva} 
  I.~Gogoladze, F.~Nasir and Q.~Shafi,
  JHEP {\bf 1311}, 173 (2013)
  [arXiv:1306.5699 [hep-ph]].


\bibitem{Kribs:2013lua} 
  G.~D.~Kribs, A.~Martin and A.~Menon,
  Phys.\ Rev.\ D {\bf 88}, 035025 (2013)
  [arXiv:1305.1313 [hep-ph]].



\bibitem{Gogoladze:2012yf} 
  I.~Gogoladze, F.~Nasir and Q.~Shafi,
  Int.\ J.\ Mod.\ Phys.\ A {\bf 28}, 1350046 (2013)
  [arXiv:1212.2593 [hep-ph]].

\bibitem{Du:2015una}
  G.~Du, T.~Li, D.~V.~Nanopoulos and S.~Raza,
  Phys.\ Rev.\ D {\bf 92}, no. 2, 025038 (2015)
  [arXiv:1502.06893 [hep-ph]].

\bibitem{Cremmer:1983bf}
  E.~Cremmer, S.~Ferrara, C.~Kounnas and D.~V.~Nanopoulos,
  Phys.\ Lett.\  B {\bf 133}, 61 (1983);
J.~R.~Ellis, A.~B.~Lahanas, D.~V.~Nanopoulos and K.~Tamvakis,
  Phys.\ Lett.\  B {\bf 134}, 429 (1984);
J.~R.~Ellis, C.~Kounnas and D.~V.~Nanopoulos,
  Nucl.\ Phys.\  B {\bf 241}, 406 (1984);
  Nucl.\ Phys.\  B {\bf 247}, 373 (1984);
A.~B.~Lahanas and D.~V.~Nanopoulos,
  Phys.\ Rept.\  {\bf 145}, 1 (1987).


\bibitem{Giudice:1988yz}
  G.~F.~Giudice and A.~Masiero,
  Phys.\ Lett.\ B {\bf 206}, 480 (1988).
\bibitem{Li:2015dil} 
  T.~Li, S.~Raza and X.~C.~Wang,
  Phys.\ Rev.\ D {\bf 93}, no. 11, 115014 (2016)
  [arXiv:1510.06851 [hep-ph]].


\bibitem{Baer:2017pba} 
  H.~Baer, V.~Barger, J.~S.~Gainer, H.~Serce and X.~Tata,
  arXiv:1708.09054 [hep-ph].
  P.~Fundira and A.~Purves,
  arXiv:1708.07835 [hep-ph].
  J.~Cao, X.~Guo, Y.~He, L.~Shang and Y.~Yue,
  arXiv:1707.09626 [hep-ph].
  B.~Zhu, F.~Staub and R.~Ding,
  Phys.\ Rev.\ D {\bf 96}, no. 3, 035038 (2017)
  [arXiv:1707.03101 [hep-ph]].
  M.~Abdughani, L.~Wu and J.~M.~Yang,
  arXiv:1705.09164 [hep-ph].
  L.~Wu,
  arXiv:1705.02534 [hep-ph].
  H.~Baer, V.~Barger, M.~Savoy, H.~Serce and X.~Tata,
  JHEP {\bf 1706}, 101 (2017)
  [arXiv:1705.01578 [hep-ph]].
  K.~J.~Bae, H.~Baer and H.~Serce,
  JCAP {\bf 1706}, no. 06, 024 (2017)
  [arXiv:1705.01134 [hep-ph]].
  T.~r.~Liang, B.~Zhu, R.~Ding and T.~Li,
  Adv.\ High Energy Phys.\  {\bf 2017}, 1585023 (2017)
  [arXiv:1704.08127 [hep-ph]].
  C.~Li, B.~Zhu and T.~Li,
  arXiv:1704.05584 [hep-ph].
  P.~S.~B.~Dev, C.~M.~Vila and W.~Rodejohann,
  Nucl.\ Phys.\ B {\bf 921}, 436 (2017)
  [arXiv:1703.00828 [hep-ph]].
  H.~Baer, V.~Barger, J.~S.~Gainer, P.~Huang, M.~Savoy, H.~Serce and X.~Tata,
  arXiv:1702.06588 [hep-ph].
  L.~Delle Rose, S.~Khalil, S.~J.~D.~King, C.~Marzo, S.~Moretti and C.~S.~Un,
  arXiv:1702.01808 [hep-ph].
  L.~Calibbi, T.~Li, A.~Mustafayev and S.~Raza,
  Phys.\ Rev.\ D {\bf 93}, no. 11, 115018 (2016)
  [arXiv:1603.06720 [hep-ph]].
  M.~van Beekveld, W.~Beenakker, S.~Caron, R.~Peeters and R.~Ruiz de Austri,
  Phys.\ Rev.\ D {\bf 96}, no. 3, 035015 (2017)
  [arXiv:1612.06333 [hep-ph]].
  M.~Peiro and S.~Robles,
  JCAP {\bf 1705}, no. 05, 010 (2017)
  [arXiv:1612.00460 [hep-ph]].


\bibitem{EENZ}
J.~R.~Ellis, K.~Enqvist, D.~V.~Nanopoulos and F.~Zwirner,
  Mod.\ Phys.\ Lett.\ A {\bf 1}, 57 (1986);



\bibitem{Barbieri:1987fn} 
  R.~Barbieri and G.~F.~Giudice,
  Nucl.\ Phys.\ B {\bf 306}, 63 (1988).


\bibitem{Baer:2012up} 
  H.~Baer, V.~Barger, P.~Huang, A.~Mustafayev and X.~Tata,
  Phys.\ Rev.\ Lett.\  {\bf 109}, 161802 (2012)
  [arXiv:1207.3343 [hep-ph]].

\bibitem{Baer:2012mv} 
  H.~Baer, V.~Barger, P.~Huang, D.~Mickelson, A.~Mustafayev and X.~Tata,
  Phys.\ Rev.\ D {\bf 87}, no. 3, 035017 (2013)
  [arXiv:1210.3019 [hep-ph]].

\bibitem{Ahmed:2016lkh} 
  W.~Ahmed, L.~Calibbi, T.~Li, A.~Mustafayev and S.~Raza,
  Phys.\ Rev.\ D {\bf 95}, no. 9, 095031 (2017)
  [arXiv:1612.07125 [hep-ph]].


\bibitem{Li:2010xr} 
  T.~Li and D.~V.~Nanopoulos,
  Phys.\ Lett.\ B {\bf 692}, 121 (2010)
  [arXiv:1002.4183 [hep-ph]].

\bibitem{Li:2014dna}
  T.~Li and S.~Raza,
  Phys.\ Rev.\ D {\bf 91} (2015) no.5,  055016
  [arXiv:1409.3930 [hep-ph]].
  
\bibitem{Li:2016ucz} 
  T.~Li, S.~Raza and K.~Wang,
  Phys.\ Rev.\ D {\bf 93}, no. 5, 055040 (2016)
  [arXiv:1601.00178 [hep-ph]].

\bibitem{Akerib:2016vxi} 
  D.~S.~Akerib {\it et al.} [LUX Collaboration],
  Phys.\ Rev.\ Lett.\  {\bf 118}, no. 2, 021303 (2017)
  [arXiv:1608.07648 [astro-ph.CO]].

\bibitem{Bennett:2006fi}
  G.~W.~Bennett {\it et al.}  [Muon G-2 Collaboration],
  Phys.\ Rev.\ D {\bf 73}, 072003 (2006);
  G.~W.~Bennett {\it et al.}  [Muon (g-2) Collaboration],
  Phys.\ Rev.\ D {\bf 80}, 052008 (2009)

\bibitem{Cheng:2012np} 
  T.~Cheng, J.~Li, T.~Li, D.~V.~Nanopoulos and C.~Tong,
  Eur.\ Phys.\ J.\ C {\bf 73}, 2322 (2013)
  [arXiv:1202.6088 [hep-ph]].

\bibitem{Balazs:2010ha} 
  C.~Balazs, T.~Li, D.~V.~Nanopoulos and F.~Wang,
  JHEP {\bf 1009}, 003 (2010)
  [arXiv:1006.5559 [hep-ph]].

 

%

\bibitem{ISAJET}
H.~Baer, F.~E.~Paige, S.~D.~Protopopescu and X.~Tata,
arXiv:hep-ph/0001086.















\bibitem{Hisano:1992jj}
J.~Hisano, H.~Murayama, and T.~Yanagida,
  { Nucl. Phys.} {\bf B402} (1993) 46.
Y.~Yamada,
{ Z. Phys.} {\bf C60} (1993) 83;
 J.~L.~Chkareuli and I.~G.~Gogoladze,
  Phys.\ Rev.\  D {\bf 58}, 055011 (1998).
\bibitem{Pierce:1996zz}
D.~M. Pierce, J.~A. Bagger, K.~T. Matchev, and R.-j. Zhang,
  { Nucl. Phys.} {\bf B491} (1997) 3.

\bibitem{Ibanez:1982fr}
L.~E. Ibanez and G.~G. Ross,
 { Phys. Lett.} {\bf B110} (1982) 215;
K.~Inoue, A.~Kakuto, H.~Komatsu and S.~Takeshita,
 { Prog. Theor. Phys.} {\bf 68}, 927 (1982)
 [Erratum-ibid.\  {\bf 70}, 330 (1983)];
L.~E. Ibanez,
{ Phys.  Lett.} {\bf B118} (1982) 73;
 J.~R. Ellis, D.~V. Nanopoulos,
and K.~Tamvakis,
  { Phys. Lett.} {\bf B121} (1983) 123;
L.~Alvarez-Gaume, J.~Polchinski, and M.~B. Wise,
{ Nucl. Phys.} {\bf B221} (1983) 495.

\bibitem{Beringer:1900zz} 
  J.~Beringer {\it et al.}  [Particle Data Group Collaboration],
  Phys.\ Rev.\ D {\bf 86}, 010001 (2012).



\bibitem{:2009ec}
    [Tevatron Electroweak Working Group and CDF Collaboration and D0 Collab],
  arXiv:0903.2503 [hep-ex].

\bibitem{bartol2} I.~Gogoladze, R.~Khalid, S.~Raza and Q.~Shafi,
  JHEP {\bf 1106} (2011) 117.
\bibitem{Belanger:2009ti}
  G.~Belanger, F.~Boudjema, A.~Pukhov and R.~K.~Singh,
  JHEP {\bf 0911}, 026 (2009);
H.~Baer, S.~Kraml, S.~Sekmen and H.~Summy,
  JHEP {\bf 0803}, 056 (2008).

\bibitem{Khachatryan:2016vau} 
  G.~Aad {\it et al.} [ATLAS and CMS Collaborations],
  JHEP {\bf 1608}, 045 (2016)
  [arXiv:1606.02266 [hep-ex]].


\bibitem{Allanach:2004rh} 
  B.~C.~Allanach, A.~Djouadi, J.~L.~Kneur, W.~Porod and P.~Slavich,
  JHEP {\bf 0409}, 044 (2004)



\bibitem{bsg} H.~Baer and M.~Brhlik, \prd{55}{1997}{4463};
H.~Baer, M.~Brhlik, D.~Castano and X.~Tata, \prd{58}{1998}{015007};
%
\bibitem{bmm} K.~Babu and C.~Kolda, \prl{84}{2000}{228};
A.~Dedes, H.~Dreiner and U.~Nierste, \prl{87}{2001}{251804};
J.~K.~Mizukoshi, X.~Tata and Y.~Wang, \prd{66}{2002}{115003}.
%


\bibitem{CMS:2014xfa}
  V.~Khachatryan {\it et al.} [CMS and LHCb Collaborations],
  Nature {\bf 522}, 68 (2015)
  [arXiv:1411.4413 [hep-ex]].



\bibitem{Amhis:2014hma}
  Y.~Amhis {\it et al.} [Heavy Flavor Averaging Group (HFAG) Collaboration],
  arXiv:1412.7515 [hep-ex].
\bibitem{Bennett:2012zja} 
  C.~L.~Bennett {\it et al.} [WMAP Collaboration],
  Astrophys.\ J.\ Suppl.\  {\bf 208}, 20 (2013)
  [arXiv:1212.5225 [astro-ph.CO]].


XENON1T
\bibitem{Aprile:2017iyp} 
  E.~Aprile {\it et al.} [XENON Collaboration],
  arXiv:1705.06655 [astro-ph.CO].




\bibitem{Aprile:2015uzo} 
  E.~Aprile {\it et al.} [XENON Collaboration],
  JCAP {\bf 1604}, no. 04, 027 (2016)
  [arXiv:1512.07501 [physics.ins-det]].




\bibitem{ATLAS:2017uun} 
  The ATLAS collaboration [ATLAS Collaboration],
  ATLAS-CONF-2017-039.

\bibitem{CMS:2017fdz} 
  CMS Collaboration [CMS Collaboration],
  CMS-PAS-SUS-16-039.

\bibitem{Baer:2014kya} 
  H.~Baer, A.~Mustafayev and X.~Tata,
  Phys.\ Rev.\ D {\bf 90}, no. 11, 115007 (2014)
  [arXiv:1409.7058 [hep-ph]].



\bibitem{Han:2013usa} 
  C.~Han, A.~Kobakhidze, N.~Liu, A.~Saavedra, L.~Wu and J.~M.~Yang,
  JHEP {\bf 1402}, 049 (2014)
  [arXiv:1310.4274 [hep-ph]].




\bibitem{ATLAS:2016uwq} 
  The ATLAS collaboration [ATLAS Collaboration],
  ATLAS-CONF-2016-096.

\bibitem{Aaboud:2017nhr} 
  M.~Aaboud {\it et al.} [ATLAS Collaboration],
  arXiv:1708.07875 [hep-ex].


\bibitem{Drees:2013wra} 
  M.~Drees, H.~Dreiner, D.~Schmeier, J.~Tattersall and J.~S.~Kim,
  Comput.\ Phys.\ Commun.\  {\bf 187}, 227 (2015)
  [arXiv:1312.2591 [hep-ph]].


\bibitem{Kim:2015wza} 
  J.~S.~Kim, D.~Schmeier, J.~Tattersall and K.~Rolbiecki,
  Comput.\ Phys.\ Commun.\  {\bf 196}, 535 (2015)
  [arXiv:1503.01123 [hep-ph]].

\bibitem{Dercks:2016npn} 
  D.~Dercks, N.~Desai, J.~S.~Kim, K.~Rolbiecki, J.~Tattersall and T.~Weber,
  arXiv:1611.09856 [hep-ph].

 \bibitem{Sjostrand:2014zea} 
   T.~Sjöstrand {\it et al.},
  Comput.\ Phys.\ Commun.\  {\bf 191}, 159 (2015)
  [arXiv:1410.3012 [hep-ph]].

\bibitem{deFavereau:2013fsa} 
  J.~de Favereau {\it et al.} [DELPHES 3 Collaboration],
  JHEP {\bf 1402}, 057 (2014)
  [arXiv:1307.6346 [hep-ex]].

\bibitem{Beenakker:1999xh} 
  W.~Beenakker, M.~Klasen, M.~Kramer, T.~Plehn, M.~Spira and P.~M.~Zerwas,
  Phys.\ Rev.\ Lett.\  {\bf 83}, 3780 (1999)
  Erratum: [Phys.\ Rev.\ Lett.\  {\bf 100}, 029901 (2008)]
  [hep-ph/9906298].

\bibitem{Alwall:2014hca} 
  J.~Alwall {\it et al.},
  JHEP {\bf 1407}, 079 (2014)
  [arXiv:1405.0301 [hep-ph]].

\bibitem{Sjostrand:2006za} 
  T.~Sjostrand, S.~Mrenna and P.~Z.~Skands,
  JHEP {\bf 0605}, 026 (2006)
  [hep-ph/0603175].


\bibitem{Cacciari:2008gp} 
  M.~Cacciari, G.~P.~Salam and G.~Soyez,
  JHEP {\bf 0804}, 063 (2008)
  [arXiv:0802.1189 [hep-ph]].


\bibitem{Xiang:2016ndq} 
  Q.~F.~Xiang, X.~J.~Bi, P.~F.~Yin and Z.~H.~Yu,
  Phys.\ Rev.\ D {\bf 94}, no. 5, 055031 (2016)
  [arXiv:1606.02149 [hep-ph]].





\end{thebibliography}
\end{document}